\PassOptionsToPackage{numbers,sort&compress}{natbib}
\documentclass[review,amsmath,amssymb,
 5p,times]{elsarticle}

\usepackage{gensymb}
\usepackage{amsmath,amssymb,color}
\usepackage{placeins}
\usepackage{dblfloatfix}
\usepackage{epstopdf}
\usepackage{color}
\usepackage{graphicx}
\usepackage{dcolumn}
\usepackage{bm}

 \usepackage[caption=false, singlelinecheck=false, justification=justified, format=plain]{subfig}

\usepackage{hyperref}
\hypersetup{
    colorlinks=true,
    citecolor=blue,
    filecolor=black,
    linkcolor=blue,
    urlcolor=black
}

\usepackage[capitalise,nameinlink]{cleveref} 
\usepackage{upgreek}

\usepackage[switch]{lineno} 

\begin{document}

\begin{frontmatter}

\title{Effects of fluid rheology and geometric disorder on the enhanced resistance of viscoelastic flows through porous media}


\author[oist]{Simon J. Haward\corref{mycorrespondingauthor} }
\cortext[mycorrespondingauthor]{Corresponding author}
\ead{simon.haward@oist.jp}

\author{Amy Q. Shen}

\address{Okinawa Institute of Science and Technology Graduate University, 1919-1 Tancha, Onna-son, Okinawa 904-0495, Japan}

\begin{abstract}
Recent works have highlighted the importance of chaotic flow fluctuations as a mechanism for the enhanced resistance observed in viscoelastic porous media flows, and have also shown that chaotic fluctuations can be either enhanced or suppressed by the structural disorder of porous media. We seek further insight into the mechanisms of flow resistance enhancement in porous media by performing pressure drop measurements and flow velocimetry on two viscoelastic fluids of contrasting rheology (one with near-constant viscosity, the other strongly shear thinning) in inertialess flow through microfluidic post arrays. Ordered hexagonal arrays have posts either ``staggered'' or ``aligned'' along the mean flow direction. Increasing disorder is applied to each configuration by randomly displacing each post within a progressively large area about its initial location. Both shear thinning and constant viscosity dilute polymer solutions show the expected increase in flow resistance for Weissenberg numbers, $\text{Wi} \gtrsim 1$. In both cases, the flow resistance enhancement increases significantly with the geometric disorder in aligned arrays, but is independent of disorder in staggered arrays. At sufficient randomisation, aligned and staggered arrays become indistinguishable, as expected. Flow velocimetry performed in selected geometries over a range of $\text{Wi}$ reveals no sign of chaotic fluctuations for the constant viscosity fluid. Instead the observation of elastic wakes between the stagnation points of the posts evokes the coil-stretch transition and implicates the extensional viscosity as the cause of the enhanced flow resistance. In contrast, the shear thinning fluid does show chaotic fluctuations for $\text{Wi}>1$, which are enhanced by disorder in aligned arrays and independent of disorder in staggered arrays (correlating with the flow resistance in this case). We further show that the first normal stress is insufficient to account for the flow resistance observed for the constant viscosity fluid, but may account for the resistance observed in the shear thinning case. Our results suggest that there may not be one universal cause for the enhanced pressure drop in viscoelastic porous media flows. Rather, the dominant mechanism may emerge depending on the specific combination of fluid rheology and geometric complexity.
\end{abstract}

\begin{keyword}
polymer solution \sep viscoelasticity \sep shear-thinning \sep microfluidics \sep porous media  \sep flow instability \sep viscosity enhancement
\end{keyword}

\end{frontmatter}

\section{Introduction}

Viscoelastic flows through porous media occur in a wide range of settings, including filtration, groundwater remediation, and enhanced oil recovery (EOR). In polymer EOR applications it is found that significantly ($\approx 30~\text{to}~50\%$) more oil can be extracted from a subterranean porous bed if a small amount, $\mathcal{O}(100~\text{parts-per-million (ppm) by weight})$, of a high molecular weight polymer is added to the base Newtonian solvent (i.e., water or brine), thus rendering the fluid viscoelastic \cite{Tyagi2025}. The reasons for this increased oil displacement have remained under investigation since the 1960's and likely involve several, potentially interrelated, factors including an anomalous enhancement of the pressure drop or flow resistance~\cite{Marshall1967,Browne2021}, a suppression of viscous fingering (or Saffman-Taylor instability)~\cite{Homsy1987,Pouplard2024}, the onset of spatio-temporal fluctuations~\cite{Clarke2015,Browne2021,Dzanic2023}, and improved dispersion and flow homogeneity~\cite{Kumar2023,Browne2023,Browne2024}. The origin of the anomalous increase in flow resistance noted by early investigators such as Marshall and Metzner \cite{Marshall1967} and Dauben and Menzie \cite{Dauben1967} for viscoelastic flows through randomly packed beds of spheres still remains under debate. 

The geometry of random porous media creates a highly complex flow field characterised by contractions and expansions, tortuous flow paths and stagnation points, resulting in a mixture of shearing and extensional kinematics. By defining a characteristic deformation rate $\dot\varepsilon$ (based simply on the ratio of flow velocity to sphere diameter) and by measuring the relaxation times $\lambda$ of their polymer solutions, Marshall and Metzner \cite{Marshall1967} showed that the flow resistance enhancement occurred at a dimensionless flow rate or Weissenberg number $\text{Wi} = \dot\varepsilon \lambda \sim 1$, implicating the onset of non-linear elastic effects in the flow. Dauben and Menzie \cite{Dauben1967} further proposed that large normal stresses, arising from the deformation of the polymer macromolecules being advected through the porespace, may explain the anomaly. James and McLaren \cite{James1975} later presented arguments suggesting that normal stresses in polymer solutions would be insufficient to account for the large flow resistance enhancements observed in experiments. Instead, by considering various features of the interconnecting porespace formed by random packings of spheres, James and McLaren \cite{James1975} concluded that the extensional kinematics of the flow field in porous media must be vital to explain the flow resistance enhancement. 

An enduring explanation for the flow resistance enhancement observed in polymeric porous media flows has been the extensional viscosity increase due to the coil-stretch transition, as predicted by de Gennes \cite{DeGennes1974} and by Hinch \cite{Hinch1974} for polymers experiencing persistent extensional flows at $\text{Wi} \gtrsim 1$. This was perhaps first proposed by Elata \textit{et al.} \cite{Elata1977}, and for some time the coil-stretch transition was the preferred explanation, being expounded on by various authors (e.g., Refs.~\citenum{Durst1981,Kulicke1984,Durst1987,Hoagland1989,Haward2003,Odell2006b,Zamani2015}). 

\begin{figure}[th!]
\begin{center}
\includegraphics[scale=0.44]{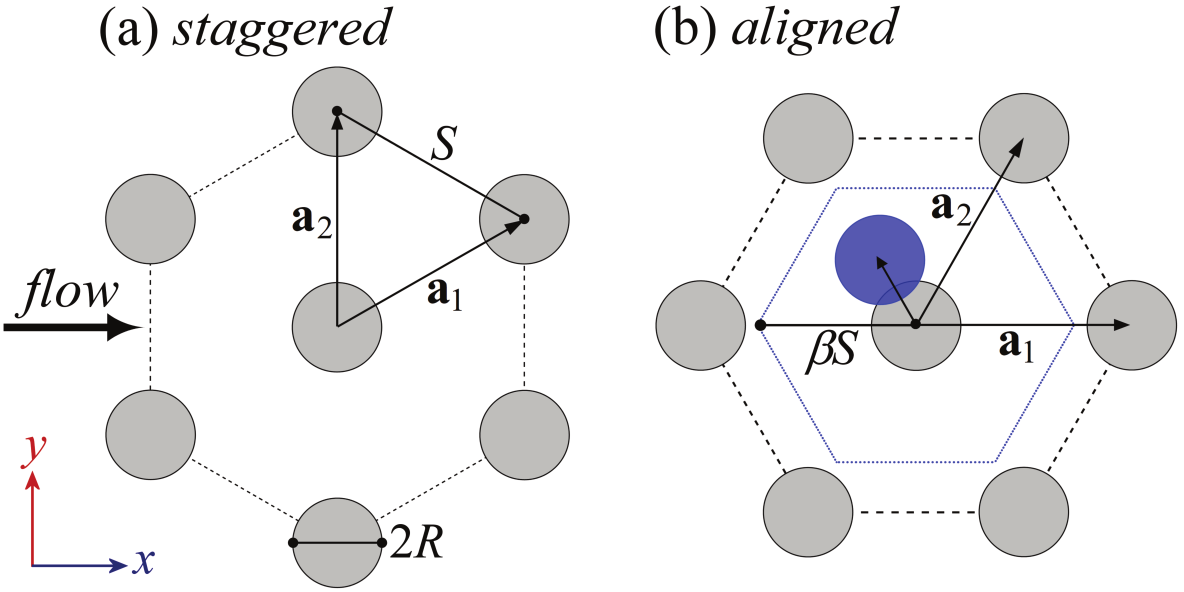}
\caption {Sketches of (a) a ``staggered'' hexagonal post array (lattice spacing $S$, post radius $R$), with the $\textbf{a}_1$ lattice vector at $30^{\circ}$ to the flow direction (i.e., along $x$), and (b) an identical ``aligned'' array with the $\textbf{a}_1$ lattice vector at $0^{\circ}$ to the flow direction. Part (b) illustrates the introduction of disorder to the arrays by random displacement of each post within a hexagon of circumradius $\beta S$ about its initial centrepoint, where $\beta$ is a variable parameter describing the degree of disorder (see Ref.~\citenum{Walkama2020}). A hypothetical new location for the central post is shown by the transparent blue circle. 
} 
\label{hex}
\end{center}
\end{figure}

Quite recently, however, a new possible explanation for the enhanced flow resistance in viscoelastic porous media flows has been receiving significant attention. Clarke \textit{et al.} \cite{Clarke2015} performed experiments in a two-dimensional (2D) microfluidic network of pores connected by narrow channels. Microscopic visualisation of the flow showed that, beyond a critical rate coincident with an increase in the apparent viscosity (or flow resistance), the flow of a viscoelastic hydrolyzed polyacrylamide (HPAA) solution became unstable and exhibited strong velocity fluctuations. In explanation, the authors related the flow resistance enhancement to the additional work required to drive the fluctuations \cite{Clarke2015}. This explanation has been broadly expanded on by Brown, Datta and coworkers~\cite{Browne2020,Browne2021} by studying viscoelastic flows through three-dimensional (3D) microfluidic models of porous media based on random packings of polydisperse glass beads. With increasing $\text{Wi}$, the authors were able to correlate the onset of flow fluctuations in increasing numbers of individual pores with the growth of the macroscopic flow resistance. The authors developed a semiquantitative model to compute the measured flow resistance as a function of the Weissenberg number based solely on the increasing viscous dissipation due to flow fluctuations, implying intriguingly that extensional viscosity was not required to explain the anomalous behaviour \cite{Browne2021}. It should be noted that Chen \textit{et al.}~\cite{Chen2024} have recently reported similar experiments performed using regular crystallographic 3D arrays of monodisperse spheres, which present significant numbers of stagnation points to the flow field. In that case, it was found that the stagnation points play a crucial role in generating flow instabilities, likely due to the large elastic stresses generated locally, and that the flow resistance enhancement exceeds that expected based on the extra viscous dissipation alone. The authors present arguments suggesting that in the ordered porous media the local extensional viscosity increase due to polymer elongation at stagnation points should be considered in order to fully account for the macroscopic flow resistance \cite{Chen2024}. Intriguingly, since extensional viscosity is not required to explain the excess pressure drop through random arrays of spheres, these results suggest that viscoelastic flow through 3D random porous media is shear dominated, as previously posited by De \textit{et al.}~\cite{De2017b}) based on numerical simulations. They also highlight the crucial role of pore geometry in controlling the pressure drop, as previously proposed and investigated by various authors (e.g.,~Refs.~\citenum{James1975,Haas1982,Durst1987,Haward2003,Odell2006b}).

The concept of viscoelastic flow fluctuations driving the pressure drop enhancement in porous media has also come under scrutiny from Ibezim \textit{et al.}~\cite{Ibezim2024}. Using random porous structures formed from sintered glass beads and several different polymer solutions, Ibezim \textit{et al.}~\cite{Ibezim2021,Ibezim2024} reported that the pressure drop enhancement occurred at much lower Weissenberg numbers than the onset of any chaotic fluctuations observed in the porespaces. Rather than correlating the nonlinear extra pressure drop to flow fluctuations therefore, they suggest that mean flow effects (i.e., polymer stretching due to the extensional flow between pores) must be responsible, at least up to the onset $\text{Wi}$ for fluctuations to set in. In addition, despite the previous resistance of James and coworkers~\cite{James1975} to considering the importance of normal stress effects, they have since shown experimentally that the first normal stress difference $N_1$ due only to the shearing kinematics in 2D ``fibrous'' porous media (regular arrays of slender posts) could fully account for the large excess pressure drops that they measured \cite{James2012,James2016}. Note also that they reported observations of viscoelastic flow instabilities in the form of asymmetries and wake structures, but claimed that all their flows were steady in time \cite{James2012}, thus the extra pressure drop in their experiments cannot be related to fluctuations.

Porous media models based on 2D arrays of posts are convenient in allowing facile visualisation of the flow \cite{Anbari2018,Browne2020}, and many authors have reported the observation of instabilities and fluctuations in viscoelastic flows through such geometries~\cite{Khomami1997,De2017,Kawale2017,Walkama2020,Haward2021}. Of most relevance to the current work, Walkama \textit{et al.}~\cite{Walkama2020} employed a microfluidic device containing an ordered hexagonal array of posts arranged with the $\textbf{a}_1$ lattice vector at $30^{\circ}$, and the posts staggered relative to the flow direction, Fig.~\ref{hex}(a). Using a viscoelastic polymer solution, they showed how the flow exhibited chaotic fluctuations for $\text{Wi} > 1$. Subsequently they introduced an increasing degree of disorder into the geometry by randomly displacing each post within an increasing area about its initial centerpoint (Fig.~\ref{hex}(b)), showing that the effect of increasing geometric disorder was to suppress the development of chaotic fluctuations \cite{Walkama2020}. Haward \textit{et al.}~\cite{Haward2021} extended the work of Walkama \textit{et al.}~\cite{Walkama2020} by starting from an ordered initial condition with the $\textbf{a}_1$ lattice vector aligned parallel to the flow direction, Fig.~\ref{hex}(b). Using a viscoelastic wormlike micellar solution, Haward \textit{et al.}~\cite{Haward2021} showed that introducing disorder to their ``aligned'' hexagonal array resulted in an enhancement of chaotic fluctuations for $\text{Wi} > 1$, in direct contrast to the effect shown by Walkama \textit{et al.} in their ``staggered'' post array geometry~\cite{Walkama2020}. The opposing results of the two authors can be reconciled by considering the opening (closing) of free paths and by the removal (introduction) of accessible stagnation points as disorder is increased in staggered (aligned) arrays~\cite{Haward2021}. Note that neither Walkama \textit{et al.}~\cite{Walkama2020} nor Haward \textit{et al.}~\cite{Haward2021} performed measurements of the flow resistance through their porous post arrays, leaving an evidential gap to be filled and possible new insights to be gained into the resistance enhancement mechanism(s). 

Accordingly, in this contribution, we further extend the work of Walkama \textit{et al.}~\cite{Walkama2020} and of Haward \textit{et al.}~\cite{Haward2021} by performing flow resistance measurements, along with time-resolved flow velocimetry, on viscoelastic flows through ordered and disordered post arrays in both staggered and aligned configurations (as per Fig.~\ref{hex}). Additional insight is provided by also examining viscoelastic fluids of contrasting rheological responses, one with an almost constant shear viscosity, and another which exhibits significant shear thinning. Even within this discrete set of experiments, we are unable to distinguish a single underlying mechanism for the flow resistance enhancement. Our results and analysis suggest that all of the mechanisms outlined above may be at play, and that the dominant mechanism in any particular scenario may depend upon the specific combination of fluid rheology and flow geometry.

\section{Materials and Methods}

\subsection{Microfluidic post arrays}

\begin{figure}
\begin{center}
\includegraphics[scale=1.2]{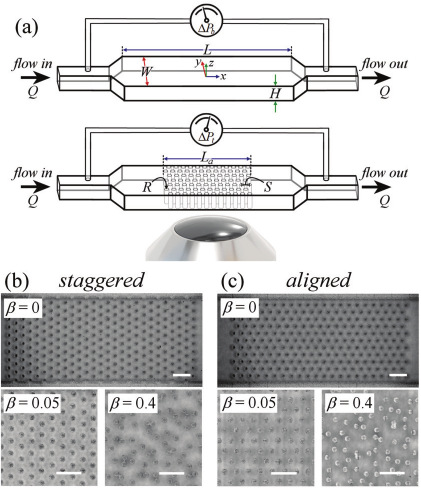}
\caption {(a) Experimental scheme (not to scale) illustrating an empty ``blank'' channel (top) and an identical channel containing an array of microposts (bottom). The coordinate system is indicated, with origin at the center of each channel. Channels have width $W=2.4~\text{mm}$, height $H=1~\text{mm}$ and length $L=25~\text{mm}$. The length of each post array is $L_a \approx 5~\text{mm}$, the radius of each post is $R=50~\upmu\text{m}$ and the lattice spacing (of ordered arrays) is $S=240~\upmu\text{m}$. Flow is driven through each channel along the $x$-direction at volumetric flow rate $Q$, while the pressure drop across each channel $\Delta P$ is measured. In post arrays, time-resolved optical interrogation of the flow field is performed using $\upmu$-PIV. Photographs of (b) ``staggered'' and (c) ``aligned'' arrays of posts with various degrees of disorder $\beta$ applied as per Ref.~[\citenum{Walkama2020}]. Scale bars: $0.5~\text{mm}$.
} 
\label{schematic}
\end{center}
\end{figure}

The microfluidic post array geometries employed in the experiments are fabricated in fused silica glass by the method of selective laser-induced etching (SLE)  using a commercial ``LightFab" 3D printer (LightFab GmbH, Germany) \cite{Gottmann2012,Meineke2016,Burshtein2019}. The microfluidic channels have width $W=2400~\upmu \text{m}$, height $H=1000~\upmu \text{m}$, and length $L=25~\text{mm}$. We employ one empty (or ``blank'') channel, along with a total of fourteen post arrays, as depicted schematically in Fig.~\ref{schematic}(a). Each post array contains around 300 posts of radius $R=50 \pm 2~\upmu \text{m}$ arranged over a length $L_a \approx 5~\text{mm}$ centered on the coordinate origin at the center of the channel. Two of the channels contain ordered hexagonal arrays of posts, lattice spacing $S=240~\upmu \text{m}$, and with the $\textbf{a}_1$ lattice vector aligned at either $30^{\circ}$ or $0^{\circ}$ to the flow direction (or $x$-axis), yielding ``staggered'' and ``aligned'' arrays, respectively (see upper images in Fig.~\ref{schematic}(b,c), respectively). Disorder is applied to each arrangement by the random displacement of each post within a radius of $\beta S$ about its initial (ordered) location, as described \cite{Walkama2020,Haward2021} (see Fig.~\ref{hex}), where $\beta$ parametrises the degree of disorder. For the ordered geometries $\beta=0$, and increasing $\beta>0$ results in increasingly disordered arrays. In this work, we use random disordering with $\beta=~$[0.05, 0.1, 0.15, 0.2, 0.3, 0.4]. Examples of ``staggered'' and ``aligned'' geometric arrangements with applied disorder are shown in the lower images of Fig.~\ref{schematic}(b,c), respectively. As can be seen in Fig.~\ref{schematic}(b,c), for $\beta=0.05$ the arrays appear to have negligible disorder, but for $\beta=0.4$ they appear significantly randomised. The post arrays used in this study all have porosity $\phi \approx 0.84$. 

\subsection{Test fluids}

\begin{figure}
\begin{center}
\includegraphics[scale=0.6]{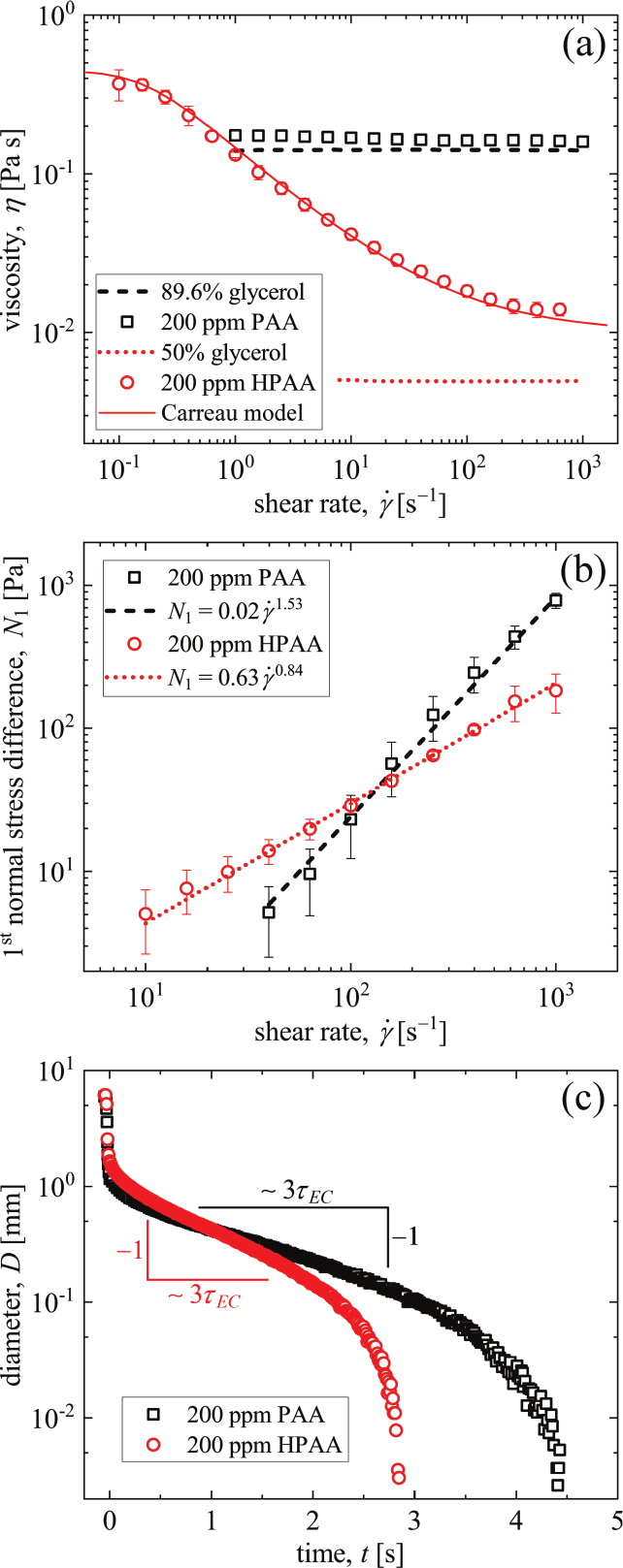}
\caption {Rheological characterisation of the polymeric test solutions at $22^{\circ}$C. (a) Flow curves of viscosity $\eta$ as a function of the shear rate $\dot\gamma$, and (b) first normal stress difference $N_1$ as a function of $\dot\gamma$ measured under steady shear using a DHR3 stress-controlled rotational rheometer (TA Instruments) fitted with a stainless steel 40~mm diameter $1^{\circ}$ cone-and-plate geometry. In (a), the shear thinning flow curve for 200~ppm HPAA is fitted with a Carreau model (Eq.~\ref{Carreau}, solid line), while respectively-coloured dashed and dotted lines indicate the solvent viscosity. Respectively-coloured dashed and dotted lines in (b) represent power-law fits to $N_1$. (c) Representative examples of the filament diameter \textit{vs.} time ($D(t)$) measured during capillary driven thinning of the polymeric fluids, from which characteristic timescales can be estimated, see main text.
} 
\label{rheol}
\end{center}
\end{figure}

Two viscoelastic polymer solutions are used in this study. The first solution is composed of a high molecular weight ($M_w \approx 5$~MDa) sample of non-ionic poly(acrylamide) (PAA, Sigma-Aldrich) dissolved to a concentration of 0.02~wt.\% (200 ppm) in a mixture of 89.6~wt.\% glycerol and 10.4~wt.\% water. The second solution is composed of an ultra-high molecular weight ($M_w \approx 18$~MDa) partially hydrolyzed poly(acrylamide) (HPAA, Polysciences Inc., degree of hydrolysis $\approx 30 \%$) dissolved to a concentration of 200~ppm in a mixture of 50~wt.\% glycerol and 50~wt.\% water. As a Newtonian reference fluid for control experiments, we employ the 89.6~wt.\% glycerol and 10.4~wt.\% water mixture with viscosity $\eta = 0.140~\text{Pa~s}$ and density $\rho = 1233~\text{kg~m}^{-3}$. The viscosities of the polymeric fluids and their solvents have been measured at $22^{\circ}$C in steady shear using a DHR3 stress-controlled rheometer (TA Instuments) fitted with a stainless steel 40~mm diameter $1^{\circ}$ cone-and-plate geometry. The results are presented in terms of viscosity $\eta$ \textit{versus} shear rate $\dot\gamma$ in Fig.~\ref{rheol}(a). As can be seen, over the tested range of shear rates ($1 \leq \dot\gamma \leq 1000~\text{s}^{-1}$), the 200~ppm PAA solution has a near-constant shear viscosity close to that of the solvent. Accordingly, we characterise its viscosity by its average value of $\eta \approx 0.165~\text{Pa~s}$. On the other hand, the 200~ppm HPAA solution is significantly shear thinning over the shear rate range $0.1 \leq \dot\gamma \leq 1000~\text{s}^{-1}$. The viscosity of this fluid is well-described by the Carreau generalised Newtonian fluid (GNF) model (shown by the fitted solid line):

\begin{equation}
\eta = \eta_{\infty} + \frac{\eta_0 - \eta_{\infty}}{[1+(\dot\gamma / \dot\gamma^{*})^{2}]^{(1-n)/2}},
\label{Carreau}
\end{equation}
where $\eta_{\infty}=0.01~\text{Pa~s}$  is the infinite-shear-rate viscosity, $\eta_0 = 0.45~\text{Pa~s}$  is the zero-shear-rate viscosity, $\dot\gamma^{*}=0.167~\text{s}^{-1}$ is the characteristic shear rate for the onset of shear thinning, and $n = 0.35$ is the ``power-law exponent'' in the shear thinning region. 

Both of the polymeric fluids exhibit nonlinear elasticity, as shown by their first normal stress difference $N_1$, measured simultaneously with the viscosity, see Fig.~\ref{rheol}(b). The data are both well described by a power-law function:

\begin{equation}
N_1(\dot\gamma) = b \dot\gamma^{m},
\label{N1}
\end{equation}
as shown by the fitted lines in Fig.~\ref{rheol}(b). For the constant viscosity 200~ppm PAA solution, $b=0.02~\text{Pa~s}^{m}$ with $m=1.53$, and for the shear thinning 200~ppm HPAA solution, $b=0.63~\text{Pa~s}^{m}$ with $m=0.84$.

Characteristic elastocapillary times ($\tau_{EC}$) for the polymeric fluids at $22^{\circ}$C are determined in a uniaxial extensional flow by measuring the diameter as a function of time ($D(t)$) of the liquid bridge generated in a capillary thinning extensional rheology device (Haake CaBER 1, Thermo Scientific) \cite{Entov1997,Anna2001b}, see Fig.~\ref{rheol}(c). The CaBER device is fitted with plates of diameter $D_0 = 6$~mm, the initial gap between the plates is 1~mm, and the plates are separated to a final gap of 6~mm by linear displacement at a rate of $0.1~\text{m~s}^{-1}$. Both test fluids display clear ``elastocapillary" thinning behaviour where $D/D_0$ decays exponentially in time $t$, and from which the elastocapillary time scale can be determined from a fit of the form: $D(t)/D_0 \sim \exp[-t/3\tau_{EC}]$. For the 200~ppm PAA solution we obtain $\tau_{EC} \approx 0.5~\text{s}$, while for the 200~ppm HPAA solution we obtain $\tau_{EC} \approx 0.36~\text{s}$. Note that due to recent debate in the literature (e.g., Ref.~\citenum{Aisling2024,Gaillard2024,Calabrese2024,Calabrese2025}), we prefer not to refer to $\tau_{EC}$ as a ``relaxation time'', since $1/\tau_{EC}$ may not always well predict the deformation rate for the onset of polymer orientation and alignment. However, it has been argued that the sensitivity of CaBER to elastic stress means that $1/\tau_{EC}$ should provide a relevant deformation rate beyond which elastic effects dominate viscous effects in extensional flows \cite{Calabrese2024,Calabrese2025}. 

\subsection{Flow Control and Dimensionless Numbers}
\label{Re,Wi,El,S}

The test fluids are driven through the microfluidic post array devices at controlled volumetric flow rates $Q$ using two neMESYS syringe pumps (low pressure, 29:1 gear ratios, Cetoni, GmbH) fitted with 10~mL Hamilton Gastight syringes. One of the pumps is used to infuse fluid into the device while the second pump withdraws fluid at an equal and opposite rate from the downstream outlet. Connections between the syringes and the microfluidic devices are made using flexible silicone tubing. The superficial flow velocity through the microchannel devices is given by $U=Q/WH$. 

The Reynolds number is used to describe the relative strength of inertial to viscous forces in the flow, and can be defined within a porous medium such as a post array by: 

\begin{equation}
 \text{Re} = \frac{2 \rho U R}{\eta(1-\phi)}, 
\label{Re}  
\end{equation}
where we take the density $\rho$ of the polymer solutions to be equal to that of the respective solvent. Based on the definition above, in experiments with the PAA solution the Reynolds number does not exceed a value of $\text{Re} \approx 0.01$. For the HPAA solution, accounting for the shear thinning viscosity, $\eta(\dot\gamma_I)$ (where $\dot\gamma_I = 2US/(S-2R)^2$ is the interstitial shear rate in the post arrays, see Sec.~\ref{eta_appHPAA}), the Reynolds number does not exceed a value of $\text{Re} \approx 0.1$. Hence, inertial effects in the flows can be neglected.

We use a characteristic Weissenberg number to describe the relative strength of elastic to viscous forces in the polymeric flows:

\begin{equation}
 \text{Wi} = \tau_{EC} \dot\varepsilon_{nom}, 
\label{Wi}  
\end{equation}
where $\dot\varepsilon_{nom} = U/\phi R$ represents a nominal deformation rate for the flow. For $ \text{Wi} \lesssim 1$, a pseudo Newtonian (viscous dominated) response of the fluid is anticipated. For $ \text{Wi} \gtrsim 1$, polymeric elastic stresses are expected to play an increasingly dominant role in controlling the dynamics (see e.g. Ref.~\cite{Haward2018,Haward2020}). 

The elasticity number $\text{El} = \text{Wi}/\text{Re}$ describes the relative strength of elastic to inertial forces in the polymeric flow. For the polymer solutions employed here, the elasticity number varies in the range $10^2 \lesssim \text{El} \lesssim 10^3$ (depending on shear thinning), indicating that (for $\text{Wi} \gtrsim 1$) the flow in the post arrays should be elasticity-dominated.

\begin{figure}[!ht!]
\begin{center}
\includegraphics[scale=0.6]{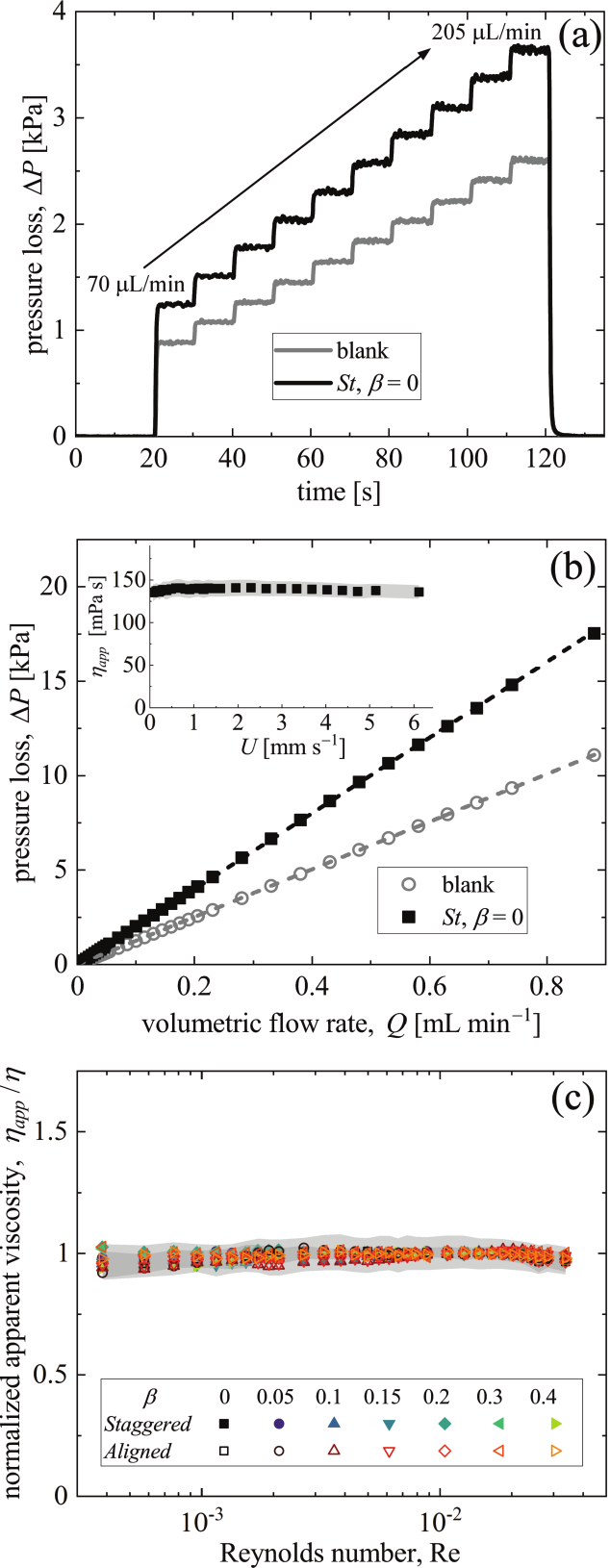}
\vspace{-0.1in}
\caption {Pressure drop measurements with the Newtonian reference fluid. (a) Example traces of the pressure drop \textit{versus} time ($\Delta P(t)$) across the blank channel and the ordered staggered array for stepwise increments of the volumetric flow rate between $70 \leq Q\leq 205~\upmu\text{L~min}^{-1}$. (b) The steady plateau pressure drop $\Delta P(Q)$ shows the expected proportionality (dashed lines). Insert shows the apparent viscosity $\eta_{app}$ in the staggered array (derived from Darcy's law, Eq.~\ref{Darcy2}) as a function of the superficial flow velocity $U$. The shaded region about the data points indicates the standard deviation over five repetitions. (c) Normalised apparent viscosity $\eta_{app}/\eta$ \textit{versus} the Reynolds number $\text{Re}$ showing a constant value $\approx 1$ for all fourteen post arrays. For better clarity, shaded error bounds are only shown for the two ordered arrays (i.e., $\beta=0$).
  } 
\label{DPNewt}
\end{center}
\end{figure}

\subsubsection{Flow Resistance Measurement}
\label{FlowResist}

We quantify the flow resistance of the various post arrays by making pressure drop measurements across the inlet and the outlet of the microfluidic chips, as indicated schematically in Fig.~\ref{schematic}(a). For this measurement, we employ a $100~\text{kPa}$ wet-wet differential pressure transducer (Omega, USA) sampling at 20~Hz. Syringe pumps are programmed to perform stepwise increments in $Q$ while the pressure drop $\Delta P(t)$ is measured continuously over time.  Fig.~\ref{DPNewt}(a) shows examples of such $\Delta P(t)$ traces made with the Newtonian reference fluid driven through ten equal flow rate increments between $70 \leq Q\leq 205~\upmu\text{L~min}^{-1}$ through both the blank channel and the ordered staggered post array. For each increment in $Q$, $\Delta P(Q)$ is determined by averaging over time in the steady-state plateau. We measure the total pressure drop $\Delta P_t(Q)$ across the chips containing post arrays, and we also measure the pressure drop $\Delta P_b(Q)$ across the blank channel. Such measurements are illustrated in Fig.~\ref{DPNewt}(b) for flow of the Newtonian reference fluid through the ordered staggered array and through the blank channel.

We perform a simple subtraction in order to remove effects of shearing on the channel walls and in the connecting upstream and downstream tubing in order to find the pressure loss due solely to the presence of the post array inside the channel:

\begin{equation}
 \Delta P_a(Q) = \Delta P_t(Q) - \Delta P_b(Q). 
\label{DPa}  
\end{equation}

The permeability of the post array is then found from Darcy's law:

\begin{equation}
 k = \frac{L_a\eta Q}{WH\Delta P_a} =  \frac{L_a\eta U}{\Delta P_a}. 
\label{Darcy1}  
\end{equation}

Rearranging Darcy's law enables an ``apparent'' viscosity to be determined from the pressure drop measurements as:

\begin{equation}
 \eta_{app} = \frac{k\Delta P_a}{L_a U}, 
\label{Darcy2}  
\end{equation}
which is shown for the ordered staggered array in the insert to Fig.~\ref{DPNewt}(b), where the shaded error bound represents the standard deviation over five repeated tests. 

We subsequently present our results in terms of a normalised apparent viscosity $\eta_{app}/\eta$, which for a Newtonian fluid in the low-Reynolds number (Darcian) regime attains a value of unity, as confirmed for all of our post arrays in Fig.~\ref{DPNewt}(c). The values for the permeability of the arrays vary slightly in the range $k=(7.3 \pm 1.3) \times10^{-10}~\text{m}^2$, with the (non-systematic) variance presumably arising from slight differences between the fabrication of each channel (e.g., variations in the precise post radius).

\subsubsection{Microparticle Image Velocimetry}
\label{PIV}

The flow of the Newtonian reference fluid and the polymeric test solutions through selected microfluidic post arrays is quantified using microparticle image velocimetry ($\upmu$-PIV, FlowMaster, LaVision GmbH \cite{Wereley2005,Wereley2010}). For these experiments, the test fluids are seeded with a low concentration ($c_p \approx 0.02$~wt\%) of 2~$\upmu$m diameter fluorescent tracer particles (PS-FluoRed, Microparticles GmbH) with excitation/emission wavelength 530/ 607~nm. The $z=0$ plane of the flow geometry is brought into focus on a microscope (SteREO Discovery V20, Zeiss) at $12 \times$ magnification, yielding a region of interest of size $\approx 2.13 \times 1.33~\text{mm}$ ($x \times y$) centred on the coordinate origin, see Fig.~\ref{schematic}(a).  For this setup, the measurement depth over which particles contribute to the determination of velocity vectors is $\delta_z \approx 230~\upmu$m (or $\approx 0.23H$) \cite{Meinhart2000}. Particle fluorescence is induced by excitation with a dual-pulsed Nd:YLF laser (wavelength of 527~nm, time separation between pulses $\Delta t$) and a high speed imaging sensor (Phantom  VEO410, Vision Research Inc.) operating in frame-straddling mode captures pairs of particle images in synchrony with the laser pulses. At each flow rate examined, the time $\Delta t$ is set so that the average displacement of particles between the two images in each pair is $\approx4$~pixels. The image capture rate and sampling duration are set depending on the fluid and the imposed flow rate. For low flow rates (or low $\text{Wi}$) image pairs are captured at 25~Hz for 8~s (200 image pairs). For higher flow rates or $\text{Wi}$ image pairs are captured at up to 40~Hz and for 40~s in order to provide a sufficient dynamic range for characterisation of any flow fluctuations. Image pairs are processed individually by a standard PIV algorithm implemented in DaVis software (LaVision GmbH), yielding 2D velocity vectors $\textbf{u} = (u_x,u_y)$ spaced at $\approx 10~\upmu$m in both $x$ and $y$.

\begin{figure}
\begin{center}
\includegraphics[scale=0.5]{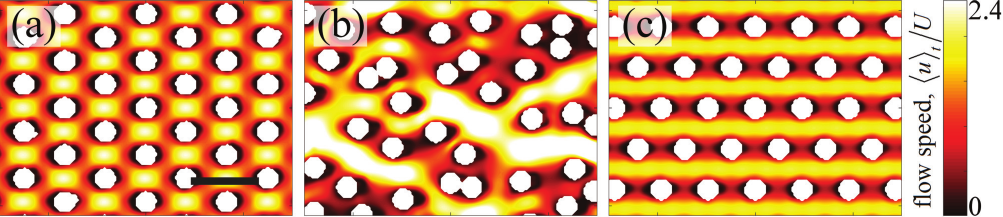}
\vspace{-0.05cm}
\caption {Normalised time-averaged flow speed $\langle u \rangle_t/U$ for the Newtonian reference fluid flowing at $\text{Re} \approx 0.02$ in a few of the post arrays: (a) Staggered, $\beta = 0$; (b) Staggered, $\beta = 0.4$; (c) Aligned, $\beta = 0$. Scale bar represents $0.2~\text{mm}$; flow is from left to right.
 } 
\label{NewtPIV}
\end{center}
\end{figure}

Fig.~\ref{NewtPIV} shows normalised time-averaged flow speed fields $\langle u \rangle_t/U$ (where $u=\lvert \textbf{u} \rvert$) measured for the Newtonian reference fluid flowing at $\text{Re} \approx 0.02$ through a few of the test geometries. The Newtonian flow remains steady in time at this low $\text{Re}$.

\section{Results and discussion}

In this section, we examine the flow of the polymeric test solutions through the various ordered and disordered microfluidic post arrays by quantifying the normalised apparent viscosity $\eta_{app}/\eta$ via pressure drop measurements and by analysing how fluctuations in the flow velocity depend on the imposed Weissenberg number. We first present the results obtained with the almost constant viscosity poly(acrylamide) solution before looking at the more shear thinning hydrolyzed poly(acrylamide) solution. 

\subsection{Constant Viscosity Poly(acrylamide) Solution}

\subsubsection{Apparent Viscosity Enhancement}

\begin{figure}[ht!]
\begin{center}
\includegraphics[scale=0.6]{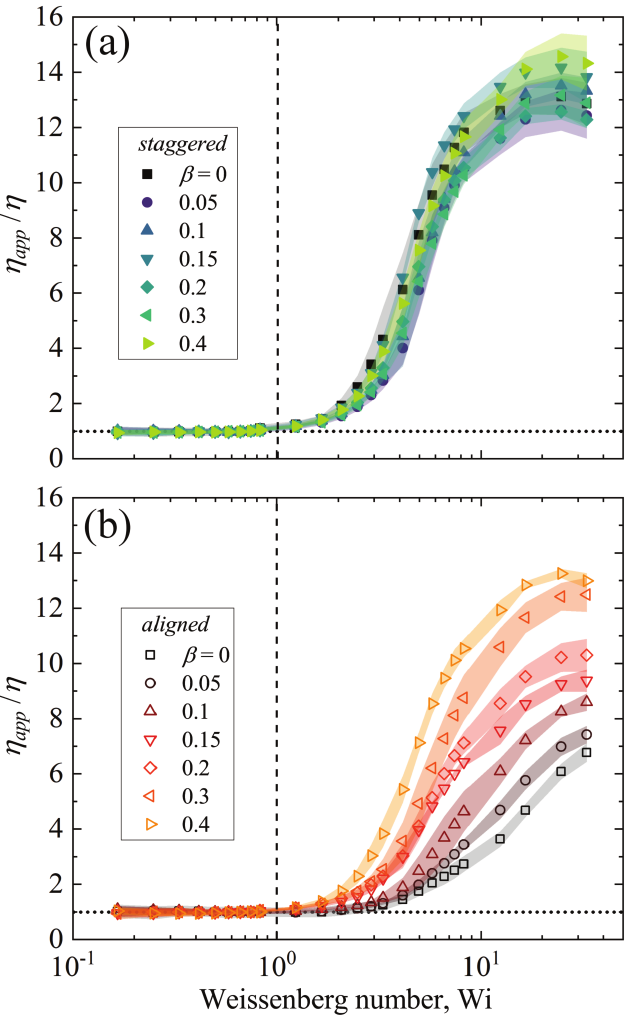}
\caption {Normalised apparent viscosity $\eta_{app}/\eta$ as a function of the Weissenberg number for flow of the 200~ppm PAA solution through (a) staggered, and (b) aligned arrays of posts. In all cases, the Reynolds number is $Re \lesssim 0.01$. Shaded error bounds represent the standard deviation over at least five repetitions. Horizontal dotted lines at $\eta_{app}/\eta = 1$ represents a pseudo-Newtonian response. Vertical dashed lines mark $\text{Wi}=1$, beyond which $\eta_{app}$ increases sharply.
 } 
\label{PAAetaapp}
\end{center}
\vspace{-0.05cm}
\end{figure}

Using the almost constant viscosity 200~ppm PAA solution ($\eta=0.165~\text{Pa~s}$) we quantify the flow resistance as a function of the Weissenberg number by following the same procedure described in Sec.~\ref{FlowResist} and illustrated for the Newtonian fluid in Fig.~\ref{DPNewt}. As shown in Fig.~\ref{PAAetaapp}, for the PAA solution at low $\text{Wi} \lesssim1$ the normalised apparent viscosity is Newtonian-like with a constant value $\eta_{app}/\eta \approx 1$. However, for all of the post arrays both staggered (Fig.~\ref{PAAetaapp}(a)) and aligned (Fig.~\ref{PAAetaapp}(b)), for $\text{Wi} \gtrsim1$, $\eta_{app}/\eta$ increases dramatically. This is a manifestation of the non-Newtonian enhancement of flow resistance classically reported in viscoelastic porous media flows, the cause of which is the subject of continuing debate. Of note here, all of the staggered arrays display similar behaviour, with $\eta_{app}/\eta$ following a similar curve regardless of the degree of disorder, see Fig.~\ref{PAAetaapp}(a). In contrast, in aligned arrays (Fig.~\ref{PAAetaapp}(b)) there is a clear systematic trend for $\eta_{app}/\eta$ to increase with disorder. Eventually, at the highest disorder of $\beta = 0.4$, the viscosity enhancement for the aligned array is comparable with that for staggered arrays.

\begin{figure}
\begin{center}
\includegraphics[scale=0.6]{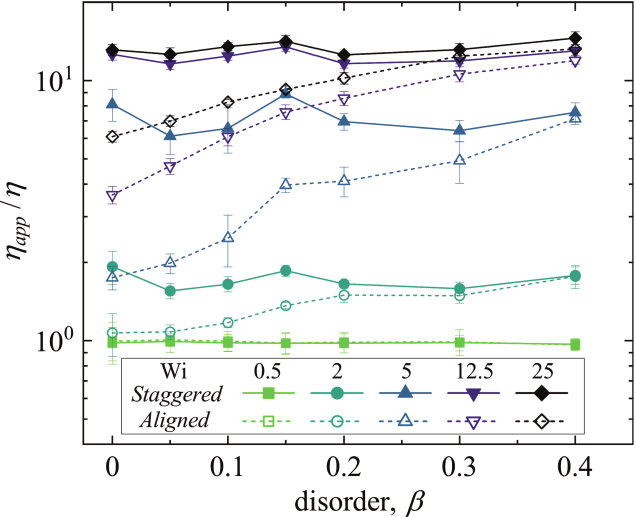}
\vspace{-0.3cm}
\caption {Normalised apparent viscosity $\eta_{app}/\eta$ as a function of the degree of disorder $\beta$ for flow of the 200~ppm PAA solution through staggered and aligned arrays of posts at several different values of the Weissenberg number.
  } 
\label{PAAetaVbeta}
\end{center}
\vspace{-0.3cm}
\end{figure}

This observation is rendered more apparent by plotting $\eta_{app}/\eta$ as a function of $\beta$ for both staggered and aligned arrays at fixed values of the Weissenberg number (see Fig.~\ref{PAAetaVbeta}). Here, for low $\text{Wi} =0.5$ flow is pseudo-Newtonian and $\eta_{app}/\eta \approx 1$ in all arrays. On the other hand, for $\text{Wi} > 1$ the flow resistance in staggered arrays increases with $\text{Wi}$ but is independent of $\beta$, while that of the aligned arrays increases progressively with $\beta$, with $\eta_{app}/\eta$ closely approaching the value observed in staggered arrays as $\beta$ reaches the maximum value of $0.4$. 

The observation that apparent viscosity is similar in both staggered and aligned arrays with $\beta=0.4$ is to be expected since at this level of disorder both types of array are significantly randomised and essentially indistinguishable. Also, the increasing trend for $\eta_{app}$ with increasing disorder in aligned arrays was expected \textit{a priori}. It was shown previously that increasing the randomisation of an initially ordered aligned array results in an increasing level of chaotic fluctuations, and it was argued that the important geometric effect of randomising an aligned array is to introduce stagnation points into the flow, thus driving large local polymer extension for $\text{Wi} > 1$ \cite{Haward2021}. Both increased fluctuations and increased polymer extension may be liable to cause an increase in $\eta_{app}$. On the other hand, it was shown by Walkama \textit{et al.}~\cite{Walkama2020} that increasing the randomisation of an initially ordered staggered array results in an decreasing level of chaotic fluctuations, and argued by Haward \textit{et al.}~\cite{Haward2021} that this would shield stagnation points from the flow. Both of these would be expected to cause a decrease in $\eta_{app}$, yet for staggered arrays we observe no trend for $\eta_{app}$ with $\beta$, which may be surprising.

To better understand how the disorder in each type of array influences the flow field and its spatiotemporal stability, we next present the results of flow velocimetry experiments performed with the 200~ppm PAA solution over a range of $\text{Wi}$ in selected staggered and aligned post arrays.

\subsubsection{Flow Velocimetry}

\begin{figure*}[ht!]
\begin{center}
\includegraphics[scale=1]{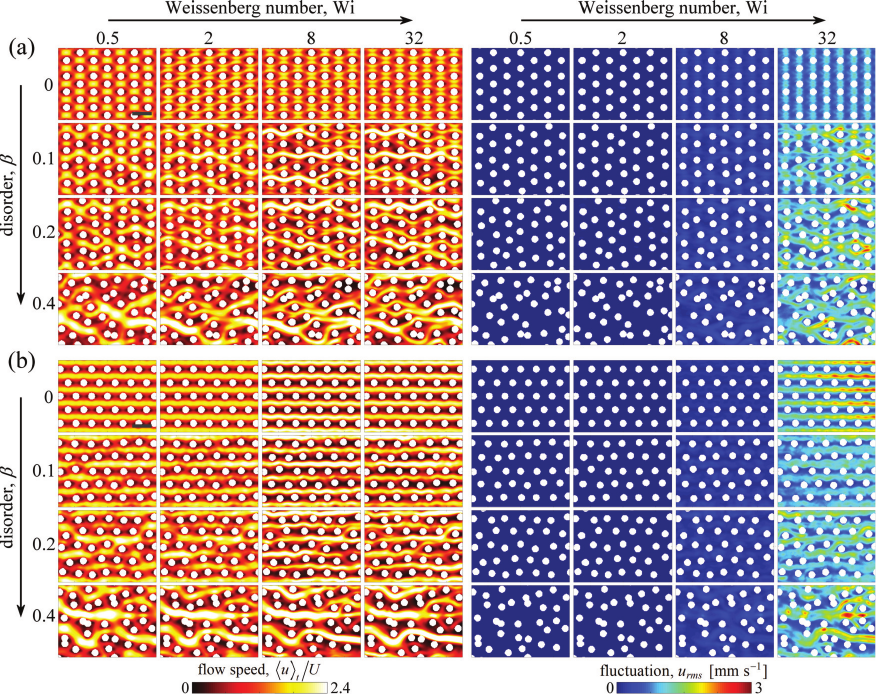}
\vspace{-0.2cm}
\caption {Normalised time-averaged flow speed (left) and root-mean-square fluctuations (right) as a function of both the disorder and the Weissenberg number for flow of the 200~ppm PAA solution in (a) staggered, and (b) aligned arrays of microposts. Scale bars represent $0.3~\text{mm}$; flow is from left to right. 
  } 
\label{PAA_PIV}
\end{center}
\end{figure*}

In Fig.~\ref{PAA_PIV}, we present the normalised time-averaged flow speed $\langle u \rangle_t/U$ (where $u=\lvert \textbf{u} \rvert$) alongside the corresponding local root-mean-square (rms) speed fluctuation $u_{rms} = \sqrt{\langle(u-\langle u \rangle_t)^2\rangle_t}$ for flow of the 200~ppm PAA solution through several staggered and aligned arrays at various values of disorder and Weissenberg number.

For the staggered arrays (Fig.~\ref{PAA_PIV}(a)) at low $\text{Wi} = 0.5$ the flow fields are essentially Newtonian-like (see Fig.~\ref{NewtPIV}(a,b)). For $\beta =0$, the flow from left to right through the array is repeatedly interrupted by the presence of the posts. From the fore-aft and lateral symmetry about each post, it is evident that the flow divides symmetrically upstream, and recombines downstream of each post. For $\text{Wi} \geq 2$, the occurrence of ``extensional wakes'' can clearly be observed downstream of each of the posts; the high speed regions downstream of each post become divided by narrow regions of lower flow speed. This type of downstream wake is characteristically formed when polymers stretch from the downstream stagnation point of cylinders or other obstacles, see e.g., Refs.~\citenum{Haward2018,Varchanis2020,Mokhtari2022,Mokhtari2024}. It is notable that the time-averaged flow patterns at $\beta =0$ are very regular even at $\text{Wi} = 32$, indicating that the flow continues to divide symmetrically around each post and suggesting the absence of instability. In contrast, when even a small degree of disorder is introduced (e.g., $\beta =0.1$) the flow patterns become spatially heterogeneous with the formation of clearly preferred flow paths. However, for staggered arrays at any level of disorder, for $\text{Wi} \geq 2$ it is possible to observe several posts in the field of view exhibiting downstream extensional wakes (i.e., extended regions of low flow speed immediately downstream of exposed stagnation points).

In contrast to staggered arrays, in aligned arrays (Fig.~\ref{PAA_PIV}(b)), at $\beta =0$ no extensional wakes can be observed in the flow fields. Independent of $\text{Wi}$, the flow is able to pass uninterrupted through the array by exploiting the gaps between the horizontal rows of aligned posts. In this case the flow bypasses the stagnation points of the posts and appears to be largely shear dominated. The situation is little changed at a low degree of disorder (e.g., $\beta=0.1$), although there is some disruption to the free-flowing paths, however for $\beta \geq 0.2$ the rearrangement of the posts is sufficient to block some of the open horizontal channels. Here, for $\text{Wi} \geq 2$, it becomes possible to locate certain posts that present extensional wakes. For $\beta =0.4$ flow fields in staggered and aligned arrays appear qualitatively similar. 

With one notable exception, apart from the inevitable measurement noise, the flow of the PAA solution appears essentially steady, with no random switching between flow paths. The only clear fluctuations beyond measurement noise are observed in the aligned array at $\beta=0$ and higher values of $\text{Wi}$. In this case, the flow speed increases and decreases along alternate flow paths, switching periodically in time (see below). 

\begin{figure}
\begin{center}
\includegraphics[scale=0.6]{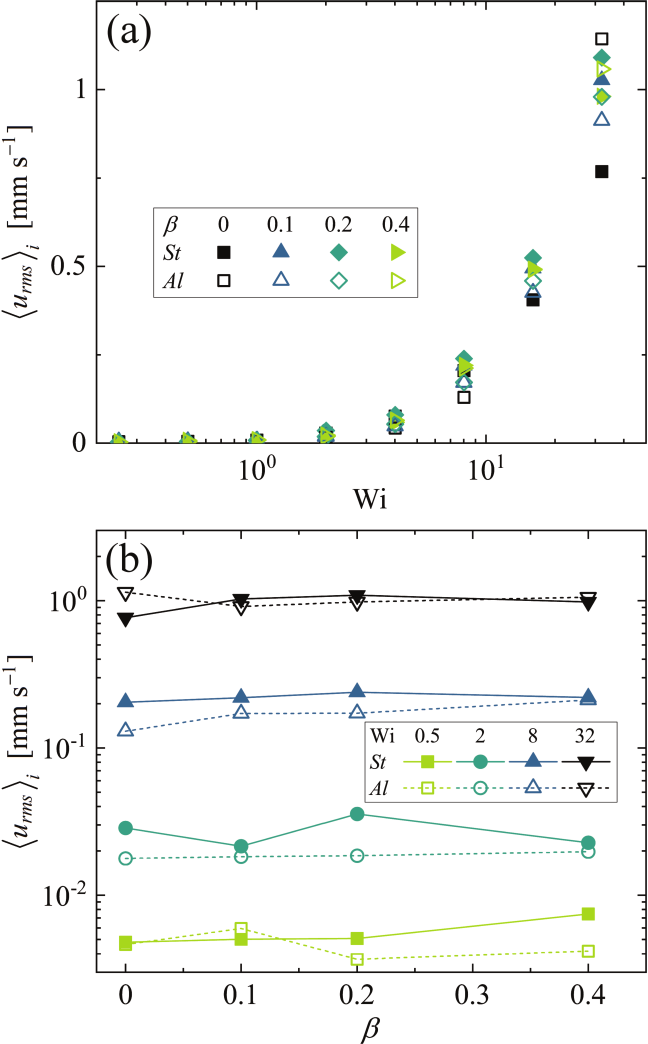}
\caption {For the 200~ppm PAA solution, spatially-averaged fluctuations $\langle u_{rms} \rangle_i$ increase with the Weissenberg number for $\text{Wi}>1$ (a), but are independent of the disorder $\beta$ in both staggered and aligned arrays (b).
 } 
\label{PAA_fluc}
\end{center}
\end{figure}

The rms fluctuations of the flow speed (right side of Fig.~\ref{PAA_PIV}) are seen to increase with the Weissenberg number. However, for a given $\text{Wi}$ there appears to be little difference depending on the array type (staggered or aligned, ordered or disordered). This is confirmed in Fig.~\ref{PAA_fluc}(a), showing how the spatially-averaged rms speed fluctuations $\langle u_{rms} \rangle_i$ increase with $\text{Wi}$ following a similar trend in all arrays. Furthermore, Fig.~\ref{PAA_fluc}(b) confirms that there is no trend of $\langle u_{rms} \rangle_i$ with $\beta$ in any of the arrays. This runs counter to expectations based on the prior works of both Walkama \textit{et al.}~\cite{Walkama2020}, who reported that fluctuations in staggered arrays decreased with increasing disorder, and of Haward \textit{et al.}~\cite{Haward2021}, who showed that fluctuations in aligned arrays increased with increasing disorder. We note here that the viscoelastic fluids used by both of those prior authors were somewhat shear thinning (in fact shear banding in the case of the wormlike micellar solution employed by Haward \textit{et al.}~\cite{Haward2021}), which contrasts to the near-constant viscosity PAA solution used currently.

\begin{figure*}[ht!]
\begin{center}
\includegraphics[scale=0.65]{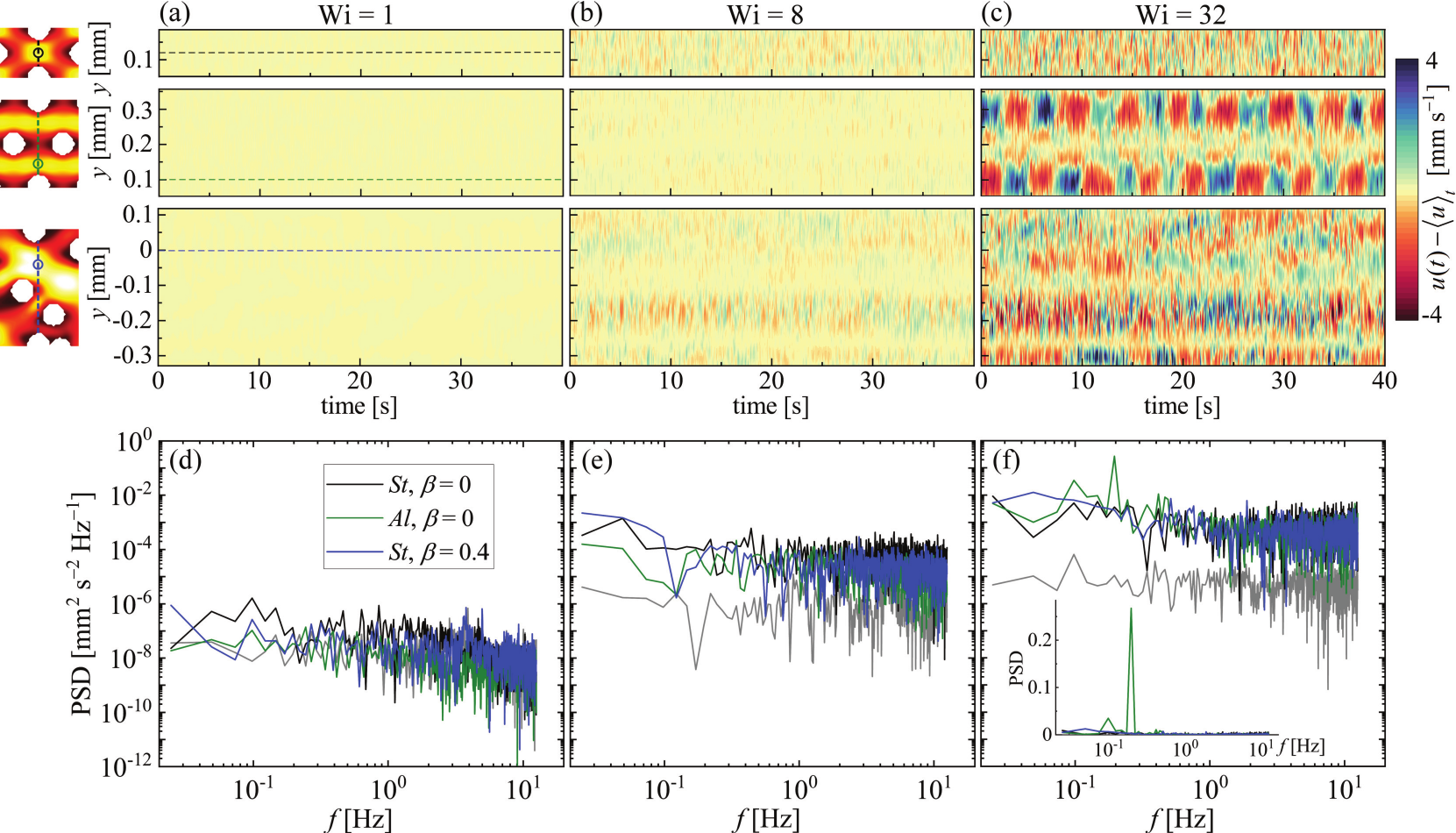}
\caption {Kymographs of the flow speed fluctuations measured for the 200~ppm PAA solution at (a) $\text{Wi}=1$, (b) $\text{Wi}=8$, and (c) $\text{Wi}=32$ in the staggered $\beta=0$ array (top row), the aligned $\beta=0$ array (middle row), and the staggered $\beta=0.4$ array (bottom row). Kymographs are made from line profiles taken between two posts located along $x=0$, as indicated by the dashed lines in the images to the left of each row. Power spectral density (PSD) of the flow speed fluctuations in each array at (d) $\text{Wi}=1$, (e) $\text{Wi}=8$, and (f) $\text{Wi}=32$ are taken along the dashed lines indicated in (a), corresponding to the points marked by the open circles in the images to the left. The gray line PSD indicates the Newtonian response in the staggered $\beta=0$ array at the volumetric flow rate respective to the Weissenberg number imposed to the polymer solution.
  } 
\label{PAA_Kym_PSD}
\end{center}
\end{figure*}

In Fig.~\ref{PAA_Kym_PSD}, we present kymographs of the flow speed fluctuations about the mean $u(t) - \langle u \rangle_t$, along with the power spectral density (PSD) of the speed fluctuations at specific locations in the flow field where  $\langle u \rangle_t$ is large. The PSDs for the 200~ppm PAA solution at various $\text{Wi}$ are compared against those from a Newtonian fluid at the corresponding superficial flow velocity. In general, the PSDs are rather flat and do not show any particular frequency dependence, indicating noise. Although at higher $\text{Wi}$ there is some amplification of noise in the PAA solution compared to the Newtonian reference fluid, PSDs do not indicate the viscoelastic response to be chaotic. As mentioned above, the only instance where we can confidently report a genuine fluctuation in the flow above the measurement noise is for the case of the aligned array with $\beta=0$, $\text{Wi}=32$. As shown by the kymograph in the middle row of Fig.~\ref{PAA_Kym_PSD}(c), this is a highly periodic fluctuation, which gives a clear peak in the PSD at a frequency $f \approx 0.2~\text{Hz}$ (Fig.~\ref{PAA_Kym_PSD}(f)). We are currently performing experiments aimed to better understand the cause and nature of this instability.

In the experiments presented so far we have seen that the flow resistance enhancement remains approximately constant in staggered arrays regardless of disorder, but increases with increasing disorder in aligned arrays. At high disorder, the flow resistance in aligned arrays matches that of the staggered arrays. The flow resistance results are partially as expected \textit{a priori} considering recent results correlating flow resistance with chaotic flow fluctuations. However, based on the work of Walkama \textit{et al.}~\cite{Walkama2020}, such a correlation would predict the flow resistance enhancement to decrease with disorder in staggered arrays. The velocimetry experiments performed in order to unravel this inconsistency have shown that in fact there are no chaotic fluctuations in any of the arrays at any value of the Weissenberg number. In fact, in the only case where an indisputably genuine flow fluctuation has been found, we observed 1) that the fluctuation is highly periodic, not chaotic, and 2) that the fluctuation occurs in the geometry yielding the smallest resistance enhancement. Certainly, for this near constant viscosity 200~ppm PAA solution, we cannot relate the enhancement in apparent viscosity through the post arrays to chaotic fluctuations in the flow field. 

As mentioned above, prior works showing how flow fluctuations depend on disorder in staggered and aligned post arrays have employed rather shear thinning viscoelastic fluids~\cite{Kawale2017,Walkama2020,Haward2021}. Indeed, studies that have related flow resistance enhancements to fluctuations in viscoelastic porous media flows have also employed fluids with significantly greater propensity for shear thinning than our 200~ppm PAA solution~\cite{Clarke2015,Browne2021}. Furthermore, low Reynolds number instabilities of viscoelastic fluids flowing around microposts have been shown to depend strongly on the shear thinning properties of the fluids~\cite{Haward2020,Varchanis2020,Hopkins2021,Haward2021b,Spyridakis2024}. With these things in mind we next proceed to examine the response of a shear thinning viscoelastic fluid, namely our 200~ppm HPAA solution, in flow through our various post arrays.  

\subsection{Shear Thinning Hydrolyzed Poly(acrylamide) Solution}

\begin{figure}[ht!]
\begin{center}
\includegraphics[scale=0.6]{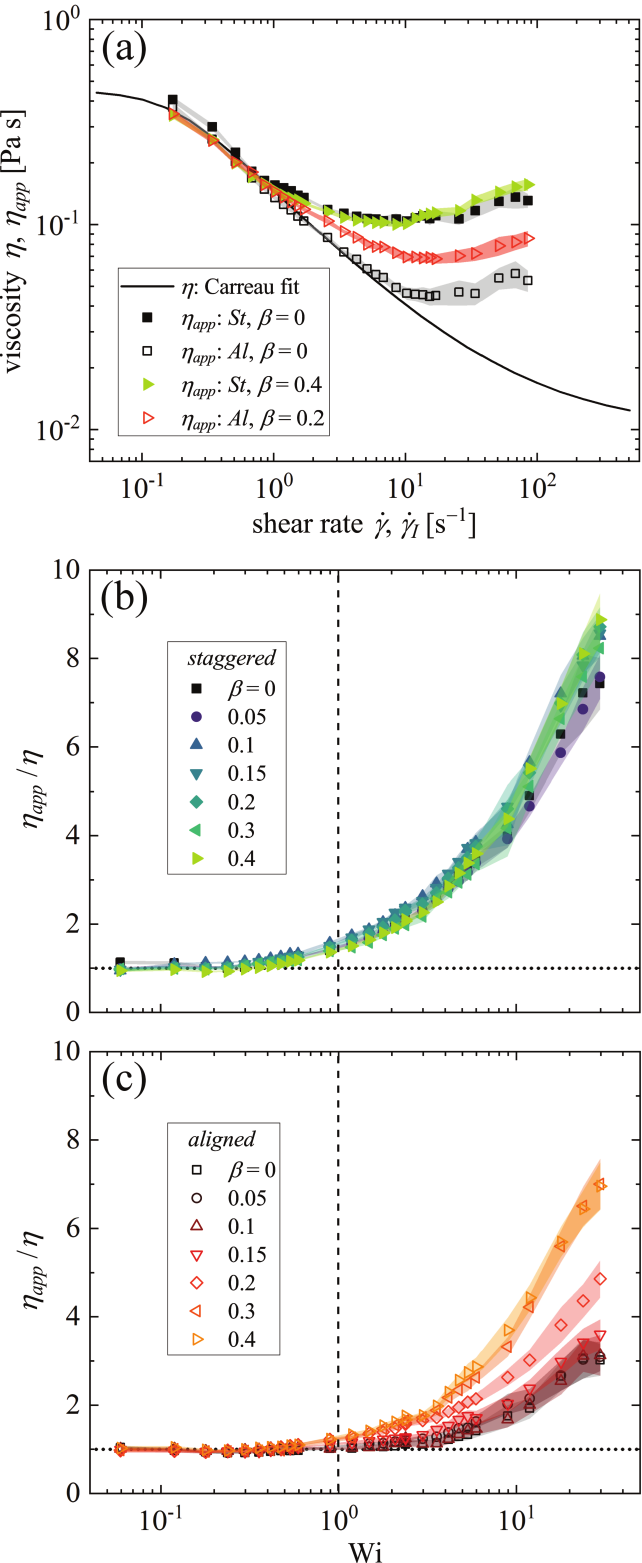}
\caption {(a) Comparison between the shear thinning flow curve $\eta(\dot\gamma)$ of the 200~ppm HPAA solution measured by rotational rheometry (solid line) and the apparent viscosity $\eta_{app}(\dot\gamma_I)$ determined from pressure drop measurements across selected post arrays and by using the Darcy equation (Eq.~\ref{Darcy2}, data points). The average interstitial shear rate in the post arrays is estimated as $\dot\gamma_I = 2US/(S-2R)^2$, which results in a fair agreement between $\eta(\dot\gamma)$ and $\eta_{app}(\dot\gamma_I)$ at low shear rates. (b, c) Normalised apparent viscosity $\eta_{app}/\eta$ as a function of the Weissenberg number for flow of the 200~ppm HPAA solution through staggered and aligned arrays of posts, respectively. In all cases, the Reynolds number is $Re \lesssim 0.1$. Shaded error bounds represent the standard deviation over at least five repetitions. Horizontal dotted lines at $\eta_{app}/\eta = 1$ represents a pseudo-Newtonian response. Vertical dashed lines mark $\text{Wi}=1$.
 } 
\label{HPAAetaapp}
\end{center}
\end{figure}
\begin{figure}[th]
\begin{center}
\includegraphics[scale=0.6]{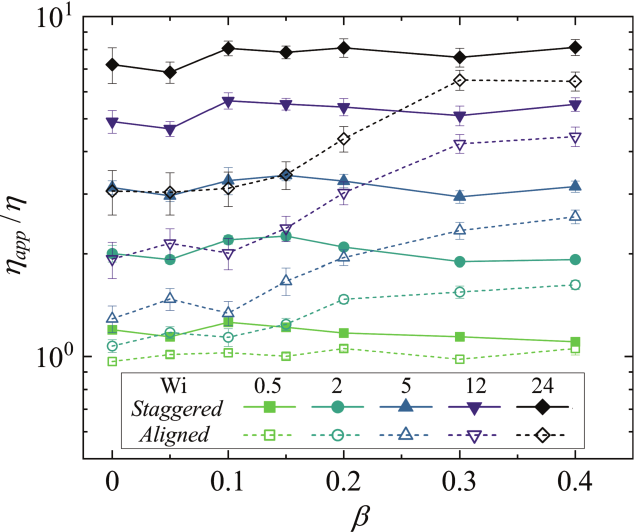}
\caption {Normalised apparent viscosity $\eta_{app}/\eta$ as a function of the degree of disorder $\beta$ for flow of the shear thinning 200~ppm HPAA solution through staggered and aligned arrays of posts at several different values of the Weissenberg number.
  } 
\label{HPAAetaVbeta}
\end{center}
\end{figure}

\subsubsection{Apparent Viscosity Enhancement}
\label{eta_appHPAA}

As described above for both the Newtonian reference fluid and the constant viscosity PAA solution, here we use pressure drop measurements across the post arrays and across the blank channel in order to determine the apparent viscosity $\eta_{app}$ of the shear thinning 200~ppm HPAA solution according to Eq.~\ref{Darcy2}. In the case of the shear thinning fluid, normalising $\eta_{app}$ by $\eta$ requires definition of an interstitial shear rate $\dot\gamma_I$ in the post arrays. For low imposed flow rates, corresponding to $\text{Wi} \lesssim 1$, we find a good agreement between our steady shear flow curve of $\eta(\dot\gamma)$ and our apparent Darcy viscosity $\eta_{app}(\dot\gamma_I)$ by setting $\dot\gamma_I = 2US/(S-2R)^2$, see Fig.~\ref{HPAAetaapp}(a). This interstitial shear rate is similar to that used by James \textit{et al.}~\cite{James2012}, defining the shear rate between the surfaces of adjacent posts in a regular array each separated by a minimum distance $S-2R$, except that here we use a smaller numerical prefactor of 2 rather than the value of 6 used by James \textit{et al}. The larger prefactor of 6 provides the maximum wall shear rate that could be expected in a post array with regular spacing $S$ and post radius $R$~\cite{James2012}. Reducing the prefactor to a value of 2 likely provides some kind of an ``average'' shear rate in the array. Conveniently, this appears to work well for both ordered and disordered arrays, as evident from the example cases shown in Fig.~\ref{HPAAetaapp}(a), and by the collapse of all the $\eta_{app}/\eta$ curves to a value of unity at low $\text{Wi} \lesssim 1$ (shown in Fig.~\ref{HPAAetaapp}(b,c)).

\begin{figure*}[ht]
\begin{center}
\includegraphics[scale=1]{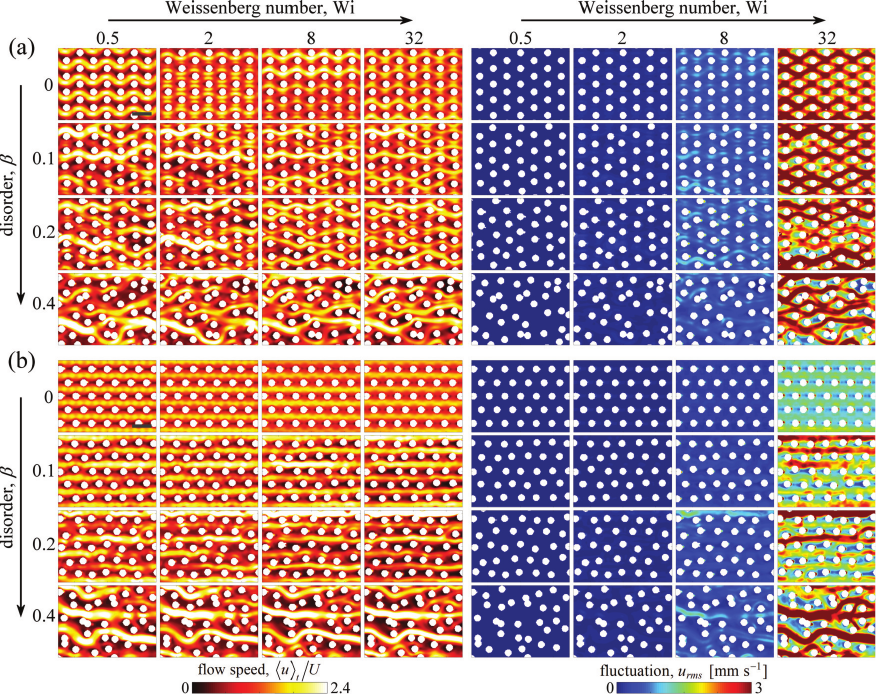}
\vspace{-0.2cm}
\caption {Normalised time-averaged flow speed (left) and root-mean-square fluctuations (right) as a function of both the disorder and the Weissenberg number for flow of the shear thinning 200~ppm HPAA solution in (a) staggered, and (b) aligned arrays of microposts. Scale bars represent $0.3~\text{mm}$; flow is from left to right. 
  } 
\label{HPAA_PIV}
\end{center}
\end{figure*}

A little unexpectedly, the curves of $\eta_{app}/\eta$ shown in Fig.~\ref{HPAAetaapp}(b,c) for the shear thinning HPAA solution follow very similar behaviour to those shown previously for the almost constant viscosity PAA solution (see Fig.~\ref{PAAetaapp}). For both the shear thinning and non-shear thinning fluids, in all tested arrays $\eta_{app}/\eta$ increases above unity for $\text{Wi} \gtrsim 1$. For staggered arrays, the curves are almost identical, and the apparent viscosity enhancement is independent of the degree of disorder (Fig.~\ref{HPAAetaapp}(b)). For aligned arrays (Fig.~\ref{HPAAetaapp}(c)), the apparent viscosity enhancement is relatively weak at low values of the disorder parameter $\beta$, but increases with increasing disorder, closely approaching the staggered case for $\beta = 0.4$. The trends described are made apparent in Fig.~\ref{HPAAetaVbeta}, showing $\eta_{app}/\eta$ \textit{versus} $\beta$ for several values of the Weissenberg number. At fixed $\text{Wi}$, $\eta_{app}/\eta$ is almost invariant with $\beta$ in staggered arrays, but (for $\text{Wi} > 1$) $\eta_{app}/\eta$ increases with $\beta$ in aligned arrays, closely approaching the line for staggered arrays at high disorder.

As before with the non-shear thinning PAA solution, we seek insight into the trends observed for the apparent viscosity enhancement in the shear thinning HPAA solution by examining the nature of the flow field in a few of the arrays.

\subsubsection{Flow Velocimetry}

The results of $\upmu$-PIV experiments performed on the shear thinning 200~ppm HPAA solution are summarised in Fig.~\ref{HPAA_PIV}. Time-averaged normalised speed fields are shown on the left hand side, and spatially resolved fields of the rms speed fluctuations $u_{rms}$ are shown on the right hand side for staggered (Fig.~\ref{HPAA_PIV}(a)) and aligned (Fig.~\ref{HPAA_PIV}(b)) arrays.

\begin{figure}[ht!]
\begin{center}
\includegraphics[scale=0.6]{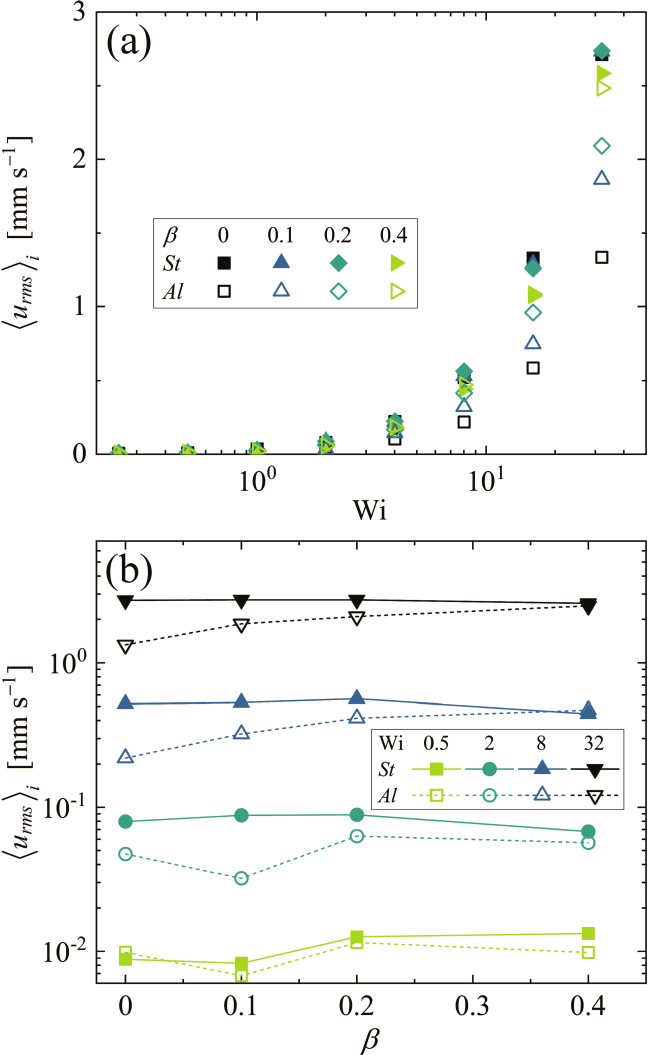}
\caption {For the shear thinning 200~ppm HPAA solution, spatially-averaged fluctuations $\langle u_{rms} \rangle  _i$ increase with the Weissenberg number for $\text{Wi}>1$ (a), similar to the less shear thinning PAA solution (see Fig.~\ref{PAA_fluc}). However, in contrast to the PAA solution, here we observe that $\langle u_{rms} \rangle  _i$ is independent of the disorder $\beta$ only in staggered arrays (b); in aligned arrays, for $\text{Wi}>1$, $\langle u_{rms} \rangle  _i$ increases with disorder, tending towards the value observed in staggered arrays.
 } 
\label{HPAA_fluc}
\end{center}
\end{figure}

For staggered arrays (Fig.~\ref{HPAA_PIV}(a)), we observe at $\beta=0$ that the flow has already destabilised at $\text{Wi} = 0.5$. Instead of the flow speed being symmetrical about each post (as seen at low $\text{Wi}$ for the PAA solution, Fig.~\ref{PAA_PIV}(a)), here the flow takes a preferential route around one side of each post, resulting in what looks like a series of parallel serpentine channels. By adopting such a configuration, the flow avoids both the leading and trailing the stagnation points of each post and becomes effectively shear dominated. At higher Weissenberg number (e.g., $\text{Wi} = 2$) a mixture of these serpentine flow paths and ``extensional wakes'' (as discussed in the context of Fig.~\ref{PAA_PIV}(a)) can be observed. In staggered arrays with higher disorder ($0.1 \leq \beta \leq 0.4$), and in aligned arrays (Fig.~\ref{HPAA_PIV}(b)) the time-averaged speed fields for the shear thinning HPAA solution are qualitatively similar to those shown previously for the non-shear thinning PAA solution (Fig.~\ref{PAA_PIV}). However, rms speed fluctuations are significantly greater in the HPAA solution than in the PAA solution (compare right side of Fig.~\ref{HPAA_PIV} with right side of Fig.~\ref{PAA_PIV}, both shown on the same colour scale). 

For the shear thinning 200~ppm HPAA solution in all the arrays, the spatially-averaged rms fluctuations $\langle u_{rms} \rangle  _i$ increase with $\text{Wi} \gtrsim 1$ (Fig.~\ref{HPAA_fluc}(a)), as also observed with the non-shear thinning PAA solution (Fig.~\ref{PAA_fluc}(a)). However, for the HPAA solution in aligned arrays we can observe a systematic increase in fluctuations with increasing $\beta$, while in staggered arrays there seems to be no particular trend (Fig.~\ref{HPAA_fluc}(a)). This observation is confirmed in Fig.~\ref{HPAA_fluc}(b), showing for fixed $\text{Wi}$ values that spatially-averaged fluctuations in staggered arrays are independent of disorder, but increase with disorder in aligned arrays (for $\text{Wi} > 1$), approaching the value observed in staggered arrays at high $\beta = 0.4$. As such, in the case of the shear thinning HPAA solution, we find some correlation between the flow resistance (or apparent viscosity, Fig.~\ref{HPAAetaVbeta}) and the degree of flow fluctuation (Fig.~\ref{HPAA_fluc}(b)). However, the result is unexpected and confusing, since the experiment performed here with the HPAA solution in staggered arrays is almost identical to that performed by Walkama \textit{et al.}~\cite{Walkama2020}, who showed a strong reduction in fluctuations as the geometric disorder increased. The geometries that we employ are a factor $\times 2$ larger than those used by Walkama \textit{et al.}~\cite{Walkama2020}, but with the same ratio of lattice spacing to post radius, thus the same porosity. Also, our HPAA test fluid is composed of the same polymer used by Walkama \textit{et al.}~\cite{Walkama2020} dissolved to almost the same concentration in a glycerol/water mixture, only we use a lower concentration of 50~wt.\% glycerol, yielding a less viscous solvent, a polymer characteristic time of roughly one-third of theirs, and slightly more shear thinning. The differences between the two experiments do not seem to be extensive, so we are unsure how to explain the stark discrepancy between our results. Note that the increasing fluctuations with increasing disorder reported here for the shear thinning HPAA solution in aligned arrays is consistent with a previous report~\cite{Haward2021}.

\begin{figure*}[ht!]
\begin{center}
\includegraphics[scale=0.65]{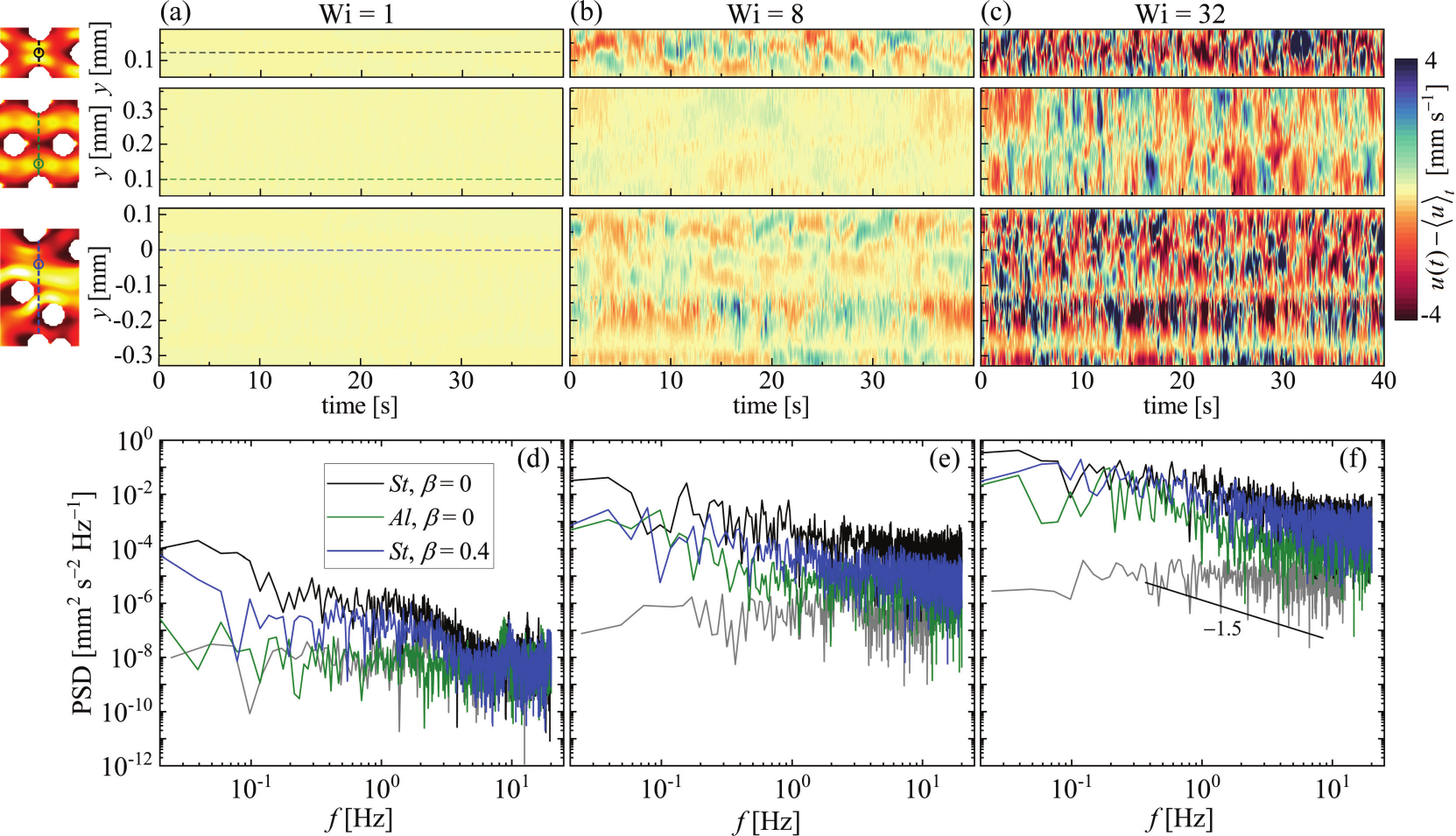}
\caption {Kymographs of the flow speed fluctuations measured for the shear thinning 200~ppm HPAA solution at (a) $\text{Wi}=1$, (b) $\text{Wi}=8$, and (c) $\text{Wi}=32$ in the staggered $\beta=0$ array (top row), the aligned $\beta=0$ array (middle row), and the staggered $\beta=0.4$ array (bottom row). Kymographs are made from line profiles taken between two posts located along $x=0$, as indicated by the dashed lines in the images to the left of each row. Power spectral density (PSD) of the flow speed fluctuations in each array at (d) $\text{Wi}=1$, (e) $\text{Wi}=8$, and (f) $\text{Wi}=32$ are taken along the dashed lines indicated in (a), corresponding to the points marked by the open circles in the images to the left. The gray line PSD indicates the Newtonian response in the staggered $\beta=0$ array at the volumetric flow rate respective to the Weissenberg number imposed to the polymer solution.
  } 
\label{HPAA_Kym_PSD}
\end{center}
\end{figure*}

For the HPAA solution, kymographs of the flow speed fluctuations about the local mean ($u(t)-\langle u \rangle_t$) and their PSDs are shown in Fig.~\ref{HPAA_Kym_PSD} (note that these are taken at the same locations in the flow field as for the PAA solution, shown previously in Fig.~\ref{PAA_Kym_PSD}, and indicated in the thumbnail images to the left of each figure). Compared with the PAA solution (Fig.~\ref{PAA_Kym_PSD}(a-c)), fluctuations in the flow speed are around twice as large in the HPAA solution (Fig.~\ref{HPAA_Kym_PSD}(a-c)), and we do not observe the periodic behaviour at high $\text{Wi}$ in the aligned array with $\beta = 0$. Comparing the PSDs between the two fluids, in the HPAA solution (Fig.~\ref{HPAA_Kym_PSD}(d-f)) the power is one to two orders of magnitude larger at low frequencies, and in contrast to the PAA solution (Fig.~\ref{PAA_Kym_PSD}(d-f)), there is a definite frequency dependence. At high Weissenberg number (e.g., $\text{Wi}=32$, Fig.~\ref{HPAA_Kym_PSD}(f)), the PSDs for the HPAA solution decay as a power law with exponent $\approx -1.5$ over at least an order of magnitude in frequency. While the slope of $-1.5$ is not steep enough to indicate elastic turbulence~\cite{Steinberg2021,Datta2022}, it does suggest the existence of chaotic features in the time-resolved flow field and is consistent with power spectrum decays reported in related post array and porous media viscoelastic flow experiments described as being chaotic~\cite{Kawale2017,Walkama2020,Browne2021}.

We have performed experiments designed to provide insight into the relationship between the enhanced apparent viscosity and chaotic flow fluctuations in viscoelastic porous media flows. Our tests have yielded some interesting, and some unexpected, results. In particular: 1) large increases in apparent viscosity have been observed in non-shear thinning PAA solution at $\text{Wi} \gtrsim 1$ in both staggered and aligned arrays, but seemingly in the absence of chaotic flow fluctuations in any of the arrays; 2) the apparent viscosity as a function of Weissenberg number is independent of disorder in staggered arrays for both constant viscosity PAA solution and shear thinning HPAA solution; 3) chaotic fluctuations are observed for the shear thinning HPAA solution but are independent of disorder, and not suppressed by disorder in staggered arrays as previously reported~\cite{Walkama2020}. In the next section of this paper we assess and discuss possible mechanisms that may help to explain our results.

\section{Assessment of Flow Thickening Mechanisms}

Following an idea suggested by Clarke \textit{et al.}~\cite{Clarke2015}, Browne and Datta~\cite{Browne2021} proposed a method to quantify the additional viscous dissipation arising due to fluctuations in viscoelastic flows at high $\text{Wi}$. They arrived at the following relation, modifying Darcy's expression for the apparent viscosity (c.f., Eqs.~\ref{Darcy1}~and~\ref{Darcy2}) with an extra term representing the extra dissipation due to the fluctuations: 

\begin{equation}
\frac{k\Delta P_a}{\eta L_a U} = \frac{\eta_{app}}{\eta} \approx 1 + \frac{k \langle \chi \rangle_{t,V}}{\eta U^2}. 
\label{Darcymod}  
\end{equation}
Here, $\langle \chi \rangle_{t,V}$ represents the time and volume-averaged rate of work done per unit volume due to fluctuations in the flow, and is quantified by:

\begin{equation}
\langle \chi \rangle_{t,V} = \eta \langle \textbf{D}'{:}\textbf{D}' \rangle_{t,V}, 
\end{equation}
where $\textbf{D}'(t) = \textbf{D}(t) - \langle \textbf{D} \rangle_{t}$, and $\textbf{D}$ is the rate-of-strain tensor. 

Using our measured 2D velocity fields, we first assess the spatiotemporally-resolved $\textbf{D}'(t)$, followed by the spatiotemporally-resolved $\textbf{D}'{:}\textbf{D}'$, from which we compute the time-averaged $\langle \chi \rangle_{t} =  \eta \langle \textbf{D}'{:}\textbf{D}' \rangle_{t}$. Finally, to estimate $\langle \chi \rangle_{t,V}$, we simply average $\langle \chi \rangle_{t}$ over the full field of view of the $\upmu$-PIV measurement. Following refs.~\citenum{Browne2021} and \citenum{Chen2024}, for each post array we fit a plot of $\langle \chi \rangle_{t,V}$ \textit{versus} $\text{Wi}$ with a power law, which is fed into Eq.~\ref{Darcymod} in order to compute $\eta_{app}/\eta$.

\begin{figure}[ht!]
\begin{center}
\includegraphics[scale=0.6]{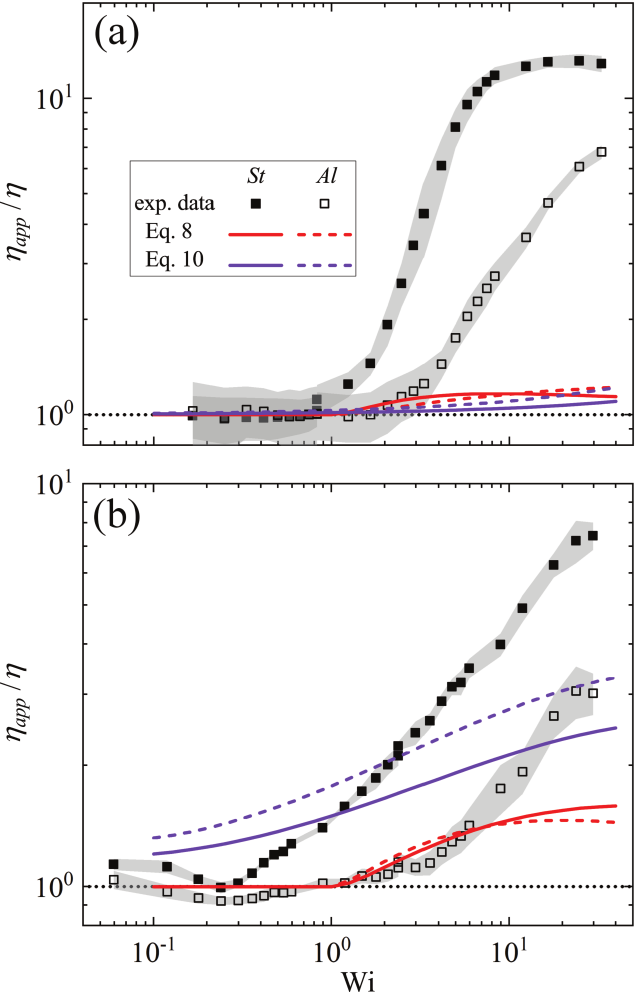}
\caption {Evaluation of Eqs.~\ref{Darcymod} and~\ref{N1stress} to assess the possible contribution of flow fluctuations and $N_1$, respectively, to the excess pressure drop measurements made with (a) the 200~ppm PAA solution, and (b) the shear thinning 200~ppm HPAA solution in a ordered ($\beta=0$) staggered and aligned post arrays. The horizontal dotted line represents the pseudo-Newtonian response.
 } 
\label{PowBal}
\end{center}
\end{figure}

Although for the non-shear thinning PAA solution we observe neither chaotic fluctuations nor any correlation between $\eta_{app}/\eta$ and $u_{rms}$, we nevertheless evaluate Eq.~\ref{Darcymod} for the two ordered post arrays and compare the results with the measured $\eta_{app}/\eta$ in Fig.~\ref{PowBal}(a) (solid and dashed red lines). Clearly, as expected, the extra viscous dissipation arising from any fluctuations (chaotic or otherwise) in the flow of the PAA solution through the post arrays is wholly insufficient to account for the large increase in the flow resistance. Even in the case of the $\beta=0$ aligned array (which shows a strong periodic fluctuation), at high $\text{Wi}$, Eq.~\ref{Darcymod} accounts for $< 5~\%$ of the apparent viscosity increase above the Darcian limit. For the shear thinning HPAA solution, we do observe potentially chaotic fluctuations and we also observe a rough correlation between $\eta_{app}/\eta$ and $u_{rms}$. However, in this case too, Eq.~\ref{Darcymod} can only partially account for the measured increase in $\eta_{app}/\eta$ (see solid and dashed red lines in Fig.~\ref{PowBal}(b)). At high $\text{Wi}$, Eq.~\ref{Darcymod} accounts for up to $\approx 20~\%$ of the apparent viscosity increase in the $\beta=0$ aligned array, but only accounts for $\approx 10~\%$ of the apparent viscosity increase in the $\beta=0$ staggered array.

Since viscous dissipation due to flow fluctuations at $\text{Wi} >1$ can (at best) only partially account for the increased flow resistance, we next consider the potential role of the first normal stress difference $N_1$, as proposed by James and coworkers~\cite{James2012,James2016}. Again we do this only for the two ordered arrays (i.e., with $\beta=0$), for which the approach described by James \textit{et al.}~\cite{James2012} is most readily transferable. As explained by James \textit{et al.} the extra elastic pressure drop through the arrays due to $N_1$ can be computed from the rotational rheometry data (i.e., Fig.~\ref{rheol}(b)) as $\Delta P_e^{N_1} = n_r \times N_1(\dot\gamma_I)$ where $n_r$ is the number of spanwise rows of posts in the array ($n_r = 43$ in the aligned array and $n_r = 27$ in the staggered array). The contribution of $N_1$ stress to the normalised apparent viscosity can thus be estimated:

\begin{equation}
\frac{\eta_{app}}{\eta} \approx 1 + \frac{k n_r N_1}{\eta L_a U^2}.
\label{N1stress}
\end{equation}

The results of this estimation are shown by the solid and dashed violet lines in Fig.~\ref{PowBal}. Clearly, in the case of the PAA solution Fig.~\ref{PowBal}(a), the extra stress due to $N_1$ is unable to account for the large increase in the observed $\eta_{app}/\eta$, providing $< 5~\%$ of the increase above the Darcian limit (similar to the contribution of flow fluctuations). However, for the HPAA solution Fig.~\ref{PowBal}(b) the contribution of $N_1$ stress is more significant, at high $\text{Wi}$ accounting for $\approx 20~\%$ of the apparent viscosity increase in the $\beta=0$ staggered array, and potentially $100~\%$ of the increase in the $\beta=0$ aligned array. Thus, while $N_1$ is clearly not a universal explanation for the enhanced resistance observed here, it may still play an important role in selected flow configurations, particularly for the shear thinning fluid.

\section{Summary and Conclusions}

In this work we have made flow resistance measurements combined with time-resolved flow velocimetry on two contrasting viscoelastic fluids flowing through two sets of microfluidic post arrays characterised by their orientation with respect to the mean flow direction and by their degree of randomisation. Given the existence of prior works showing how the randomisation of post arrays can affect chaotic viscoelastic flow fluctuations, our main objective was to test the proposed link between chaotic fluctuations and the long-debated anomalous increase in flow resistance through porous media.

Surprisingly to us, our first experiments with a near-constant viscosity PAA solution showed no chaotic fluctuations at all in either of our two sets of post geometries (staggered or aligned, with or without disorder). However, in both staggered and aligned arrays we observed large increases in the flow resistance above a flow rate corresponding to a Weissenberg number $\text{Wi} \approx 1$. In staggered arrays, the large flow resistance enhancement was independent of disorder, but in aligned arrays the enhancement increased with disorder, matching that of staggered arrays as disorder became significant. Clearly, in this case we see no relation between chaotic fluctuations and the increase in flow resistance.

For our shear thinning test fluid, we did observe flow fluctuations that could be described as chaotic, especially as the flow rate was increased to provide  $\text{Wi} \gtrsim 2$. For staggered arrays, fluctuations observed at a given value of $\text{Wi}$ were independent of the geometric disorder, but in aligned arrays the fluctuations increased with geometric disorder, in line with a previous report~\cite{Haward2021}. In this case we observed a rough correlation between the fluctuations and the flow resistance increase, which was also independent of disorder in staggered arrays, but increased with disorder in aligned arrays. 

The observation of fluctuations in the shear thinning fluid, but not in the constant viscosity fluid, may be explained by considering the elastic instability criterion of Pakdel and McKinley~\cite{Pakdel1996,McKinley1996}, which predicts the onset of instability when:

\begin{equation}
\frac{\lambda U}{\mathcal{R}}  \frac{N_1}{\eta \dot\gamma} > M_{crit}^2 
,
\label{pakmac}
\end{equation}
where the relaxation time $\lambda$ can be taken equal to $\tau_{EC}$, and $\mathcal{R}$ is a characteristic radius of curvature of a streamline. Since the values of $\tau_{EC}$ are similar for the two viscoelastic test fluids, then for a given flow rate in a given geometry, the magnitude of $M^2$ is primarily determined by the second term in Eq.~\ref{pakmac}, $N_1/\eta \dot\gamma$. For the HPAA solution the flow becomes unsteady for Weissenberg numbers $2 \lesssim \text{Wi} \lesssim 4$, corresponding to interstitial shear rates $5 \lesssim \dot\gamma_I \lesssim 10~\text{s}^{-1}$.  Over this range of shear rate, by extrapolation of our $N_1$ data (Fig.~\ref{rheol}(b)), we obtain values of $8 \lesssim N_1/\eta \dot\gamma \lesssim 10$ for the HPAA solution but only of $0.2 \lesssim N_1/\eta \dot\gamma \lesssim 0.4$ for the PAA solution. Despite the fact that $N_1$ increases more rapidly with shear rate for the PAA solution than for the HPAA solution (Fig.~\ref{rheol}(b)), the non-shear thinning viscosity of the PAA solution means that $N_1/\eta \dot\gamma$ remains relatively low even as $\text{Wi}$ becomes large. Even at the highest tested $\text{Wi}=32$ (corresponding to $\dot\gamma_I \approx 85~\text{s}^{-1}$) we find $N_1/\eta \dot\gamma \approx 1.2$, suggesting that perhaps we only approached the borderline threshold for instability in this case. This analysis illustrates how the fluid rheology may influence the stability of viscoelastic porous media flows. A more systematic study of rheological effects in key ordered and disordered post arrays, perhaps along the lines of Refs.~\citenum{Haward2020,Varchanis2020,Yokokoji2024}, would aid a deeper understanding.

We evaluated a recently-proposed power-balance incorporating the extra viscous dissipation due to flow fluctuations into Darcy's law~\cite{Browne2021}. As expected, for the non-shear thinning fluid, since there are no strong fluctuations, this approach is unable to account for the large increase observed in the flow resistance. Although it does a slightly better job for the shear thinning fluid, the model is also unable to fully explain the resistance increase in this case. Since flow fluctuations cannot account for our measured increases in flow resistance, we also evaluated the role of $N_1$ stresses~\cite{James2012,James2016}, again showing that these are generally insufficient to explain our excess pressure drop data (except perhaps in the specific case of the shear thinning fluid flowing through the aligned post array with no disorder). 

One of the key prior works motivating the current work showed that viscoelastic flow fluctuations are suppressed by applying geometric disorder to a staggered configuration of posts~\cite{Walkama2020}. Unexpectedly, we did not observe this phenomenon with either of our two viscoelastic test fluids (one shear thinning, and the other with near constant shear viscosity). While we are unsure how to explain this discrepancy, it appears that fluctuation suppression of the type reported in Ref.~\citenum{Walkama2020} may be dependent on a rather specific combination of geometric and rheological parameters. Despite the slightly negative impact of this discrepancy on our initial experimental plan, we remain able to make several statements regarding the possible mechanism(s) of flow resistance enhancement in viscoelastic porous media flows. Firstly, we conclude that chaotic flow fluctuations are not a necessary requirement for the generation of large enhancements in the  flow resistance, and in general chaotic fluctuations are insufficient to account for observed resistance increases.  However, in some cases they may play a significant contributing, or even dominant role, as clearly shown in prior works~\cite{Browne2021,Chen2024}. We can also make similar statements about the role of $N_1$ stresses~\cite{James2012}.

For the near-constant viscosity PAA solution in particular, neither viscous dissipation from flow fluctuations nor first normal stress differences appear to play a significant role in driving the flow resistance enhancement. However, the extensional wakes visible downstream of stagnation points in the staggered PAA flow fields (Fig.~\ref{PAA_PIV}(a)) provide direct evidence that polymers accumulate strain under the locally sustained extensional kinematics, suggesting that an increase in extensional viscosity due to the coil-stretch transition is the dominant contributor. In the ordered aligned array, where the flow bypasses stagnation points entirely through horizontal gaps between rows of posts (Fig.~\ref{PAA_PIV}(b)), extensional kinematics are instead generated in the contractions between rows; extensional viscosity remains the natural candidate mechanism, though the stagnation-point picture does not apply directly in this case (note that even in the ordered aligned array, there are, in fact, a few trailing stagnation points associated with the ultimate downstream spanwise row of posts). A natural next step would be to incorporate an apparent extensional viscosity, evaluated from the accumulated Hencky strain in the pore-scale flow field combined with transient extensional rheology measurements, directly into a modified power balance. Such an approach would provide a quantitative test of the extensional viscosity mechanism and would extend the framework of Browne and Datta~\cite{Browne2021} to ordered geometries in which stagnation-point-driven or constriction-driven extensional flow dominates over chaotic fluctuations and $N_1$ stress.

In summary, our results demonstrate that enhanced flow resistance in viscoelastic porous media cannot be universally attributed to chaotic flow fluctuations. While fluctuations may contribute in some cases (as in our shear thinning fluid), they are virtually absent in our near-constant viscosity polymer solution despite the even larger resistance enhancement. This suggests that various mechanisms, including extensional viscosity, normal stresses, and flow fluctuations, play a role and the dominant mechanism may depend on the combination of fluid rheology and pore geometry. Future work should focus on incorporating extensional effects more directly into predictive models for the pressure drop in viscoelastic porous media flow.

\section*{Acknowledgements}

We gratefully acknowledge the support of the Okinawa Institute of Science and Technology Graduate University (OIST) with subsidy funding from the Cabinet Office, Government of Japan. We also acknowledge Kakenhi funding from the Japan Society for the Promotion of Science (Grant Nos. 24K07332, and 24K00810). We are indebted to Prof.~Sujit Datta (Caltech) and to Dr.~Vincenzo Calabrese (Polymat, University of the Basque Country) for helpful discussions and insightful comments on the manuscript.



\end{document}